\documentclass[pra,floatfix,twocolumn,showpacs,amsmath,amssymb]{revtex4}
\usepackage{graphicx}
\usepackage{dcolumn}
\usepackage{bm}
\usepackage{color}
\usepackage{ulem}

\begin{document}
\title{
De Haas-van Alphen oscillations near the Lifshitz transition from two electron pockets to one electron pocket in the two-dimensional Dirac fermion systems 
}

\author
{Keita Kishigi}
\affiliation{Faculty of Education, Kumamoto University, Kurokami 2-40-1, 
Kumamoto, 860-8555, Japan}

\author{Yasumasa Hasegawa}
\affiliation{Department of Material Science, 
Graduate School of Material Science, 
University of Hyogo, Hyogo, 678-1297, Japan}

\date{\today}

\begin{abstract}

We theoretically study the de Haas-van Alphen (dHvA) oscillations in the 
system with changing the topology of the Fermi surface (the Lifshitz transition)
by electron dopings. We employ the two-dimensional tight binding model 
for $\alpha$-(BEDT-TTF)$_2$I$_3$ under pressure which has two Dirac points in the first Brillouin zone. 
When this system is slightly doped, there exists two closed Fermi surfaces with the same area and the dHvA oscillations become saw-tooth pattern or inversed saw-tooth pattern for both cases of fixed electron filling ($\nu$) or fixed chemical potential ($\mu$) with respect to the magnetic field, respectively. By increasing dopings, the system approaches the Lifshitz transition, where two closed Fermi surfaces are close each other. Then, we find that the pattern of the dHvA oscillations changes. A jump of the magnetization appears at the center of the fundamental period and its magnitude increases in the case of the fixed electron filling, while a jump is separated into a pair of jumps and its separation becomes large in the case of the fixed chemical potential. This is due to the lifting of double degeneracy in the Landau levels. Since this lifting is seen in the two-dimensional Dirac fermion system with two Dirac points, the obtained results in this study can be applied to not only $\alpha$-(BEDT-TTF)$_2$I$_3$ but also other materials with closely located Dirac points such as graphene under the uniaxial strain, in black phosphorus, twisted bilayer graphene, and so on. 
\end{abstract}

\date{\today}

\maketitle

\section{Introduction}


The kinetic energy of electrons perpendicular to an applied magnetic field ($H$) is quantized as Landau levels\cite{shoenberg,Landau,Onsager}. As a result, the magnetization oscillates as a function of
the inverse of the magnetic field ($1/H$) at low temperatures,
which is called the de Haas van Alphen oscillations\cite{shoenberg,dHvA} and gives an important information of the cross-sectional area of the Fermi surface. 
However, when the system undergoes the Lifshitz transition\cite{Lifshitz}, the topology and the area of the Fermi surface change. Therefore, it is expected that interesting phenomena occur in dHvA oscillations near the Lifshitz transition.

In this paper we study the dHvA oscillations numerically by using the tight-binding model with the Peierls phases\cite{machida,kishigi,sandu}. In this approach the field-induced quantum tunneling is studied without the semiclassical approach of magnetic breakdown\cite{shoenberg}.


In a previous study\cite{KH2017}, we have calculated the energies under the magnetic field in the tight-binding model for $\alpha$-(BEDT-TTF)$_2$I$_3$, which is known as one of quasi-two-dimensional Dirac fermions systems, where there are one hole pocket and one or two electron pocket(s). 
We have shown the lifting of double degenerated electron pocket's Landau levels, which is caused by the field-induced quantum tunneling. A similar lifting has been shown in the graphene with the
anisotropic transfer integral of two-dimensional Dirac fermions system\cite{HK2006} and in the simple model with two electron pockets\cite{Montambaux2009}. 
These liftings are seen near the Lifshitz transition. 

Recently, we have studied\cite{KH2019} the dHvA oscillations near the Lifshitz transition (i.e., in the case when the lifting of the Landau levels occurs) in the two-dimensional compensated metallic system with one hole pocket and one or two electron pockets, where one electron pocket is transformed into two electron pockets by applying the uniaxial pressure. As a result, we have found that the Fourier transform intensities (FTIs) for the frequencies corresponding to the 3/2 and 5/2 times area of a hole pocket are enhanced at the Lifshitz transition\cite{KH2019}. We have explained that the enhancement of the 3/2 times frequency is caused by the commensurate separation of doubly degenerated Landau levels with the phase factor ($\gamma_e\simeq 0$). 

\begin{figure}[bt]
\begin{center}
\begin{flushleft} \hspace{0.5cm}(a) \end{flushleft}\vspace{-0.5cm}
\includegraphics[width=0.46\textwidth]{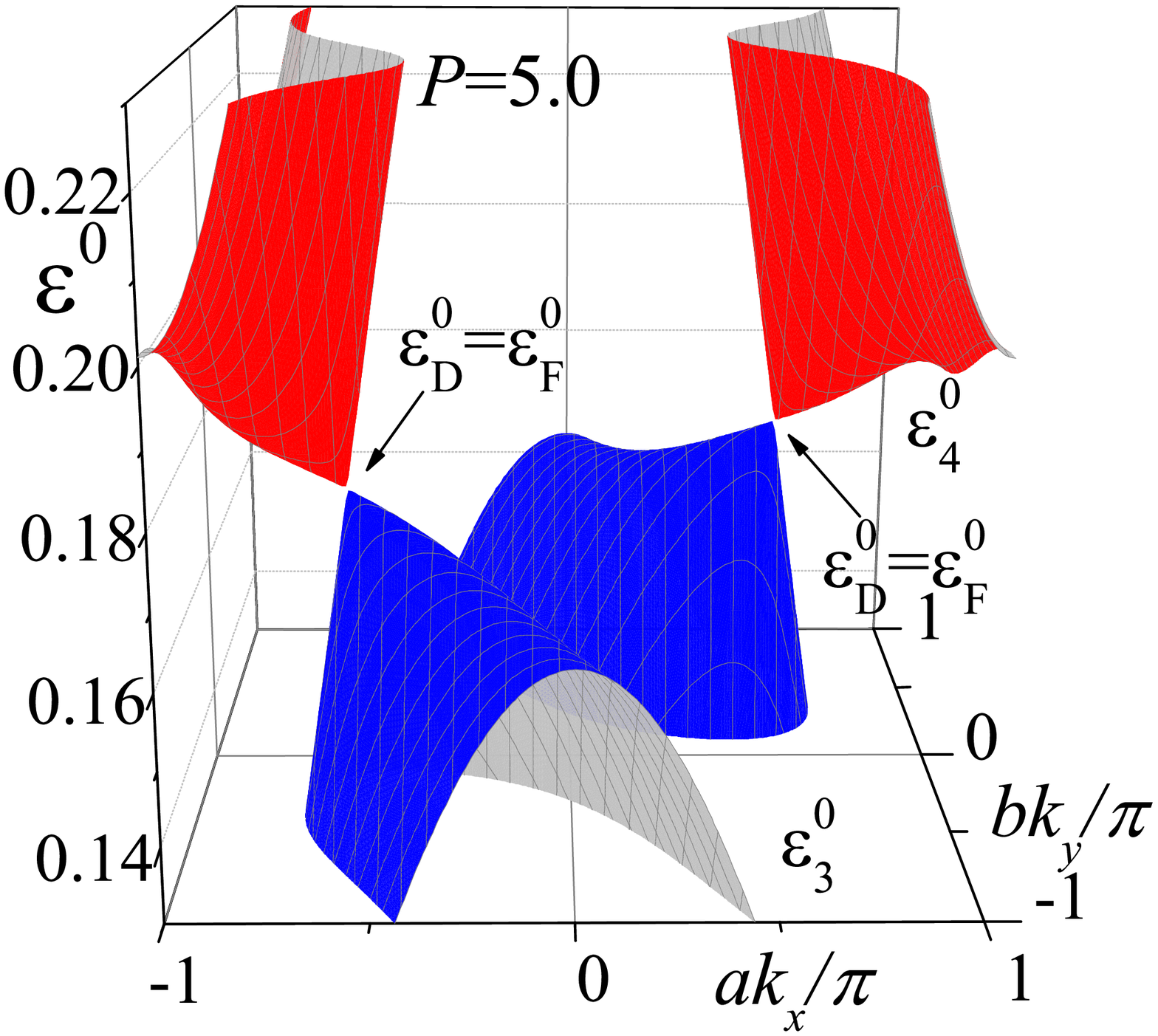}\vspace{-0.5cm}
\begin{flushleft} \hspace{0.5cm}(b) \end{flushleft}\vspace{-0.5cm}
\includegraphics[width=0.46\textwidth]{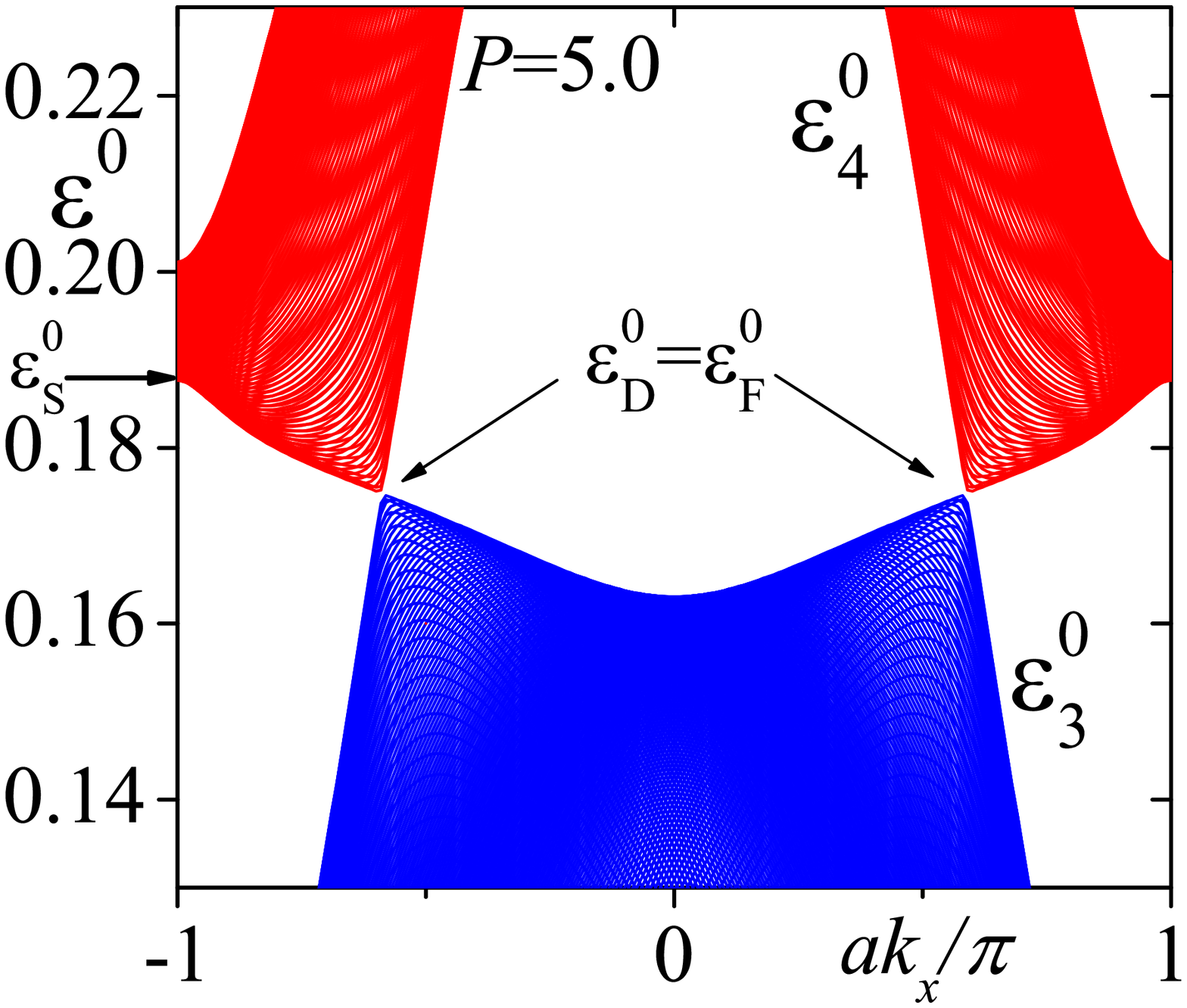}\vspace{-0.5cm}
\begin{flushleft} \hspace{0.5cm}(c) \end{flushleft}\vspace{-0.5cm}
\includegraphics[width=0.46\textwidth]{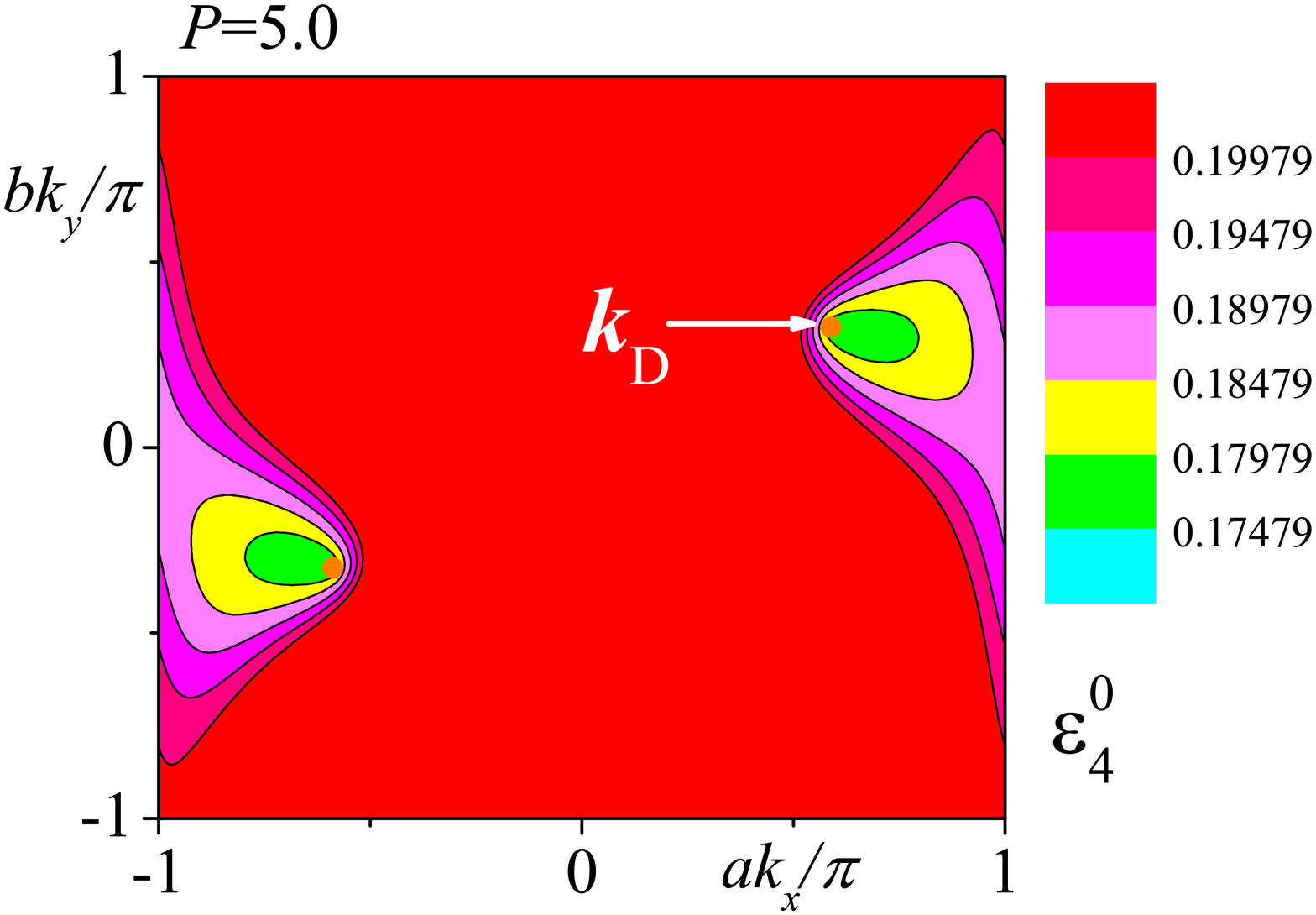}\vspace{-0.5cm}
\end{center}
\caption{
(Color online) 
(a)  The third and fourth energy bands ($\varepsilon_{\rm 3}^0$ 
and $\varepsilon_{\rm 4}^0$) at $P=5.0$. (b) is a figure of (a) from a distant view point along the $k_y$ axis. $\varepsilon_{\rm s}^0$ is the energy of the saddle point, $\varepsilon_{\rm D}^0$ is the energy at the Dirac points and $\varepsilon^0_{{\rm F}}$ is the Fermi energy at 3/4-filling, where $\varepsilon_{\rm D}^0=\varepsilon_{\rm F}^0\simeq 0.17479$. 
 (c) Contour plots of fourth band. 
}
\label{fig2}
\end{figure}

\begin{figure}[bt]
\begin{center}
\begin{flushleft} \hspace{0.5cm}(a) \end{flushleft}\vspace{-0.5cm} 
\includegraphics[width=0.44\textwidth]{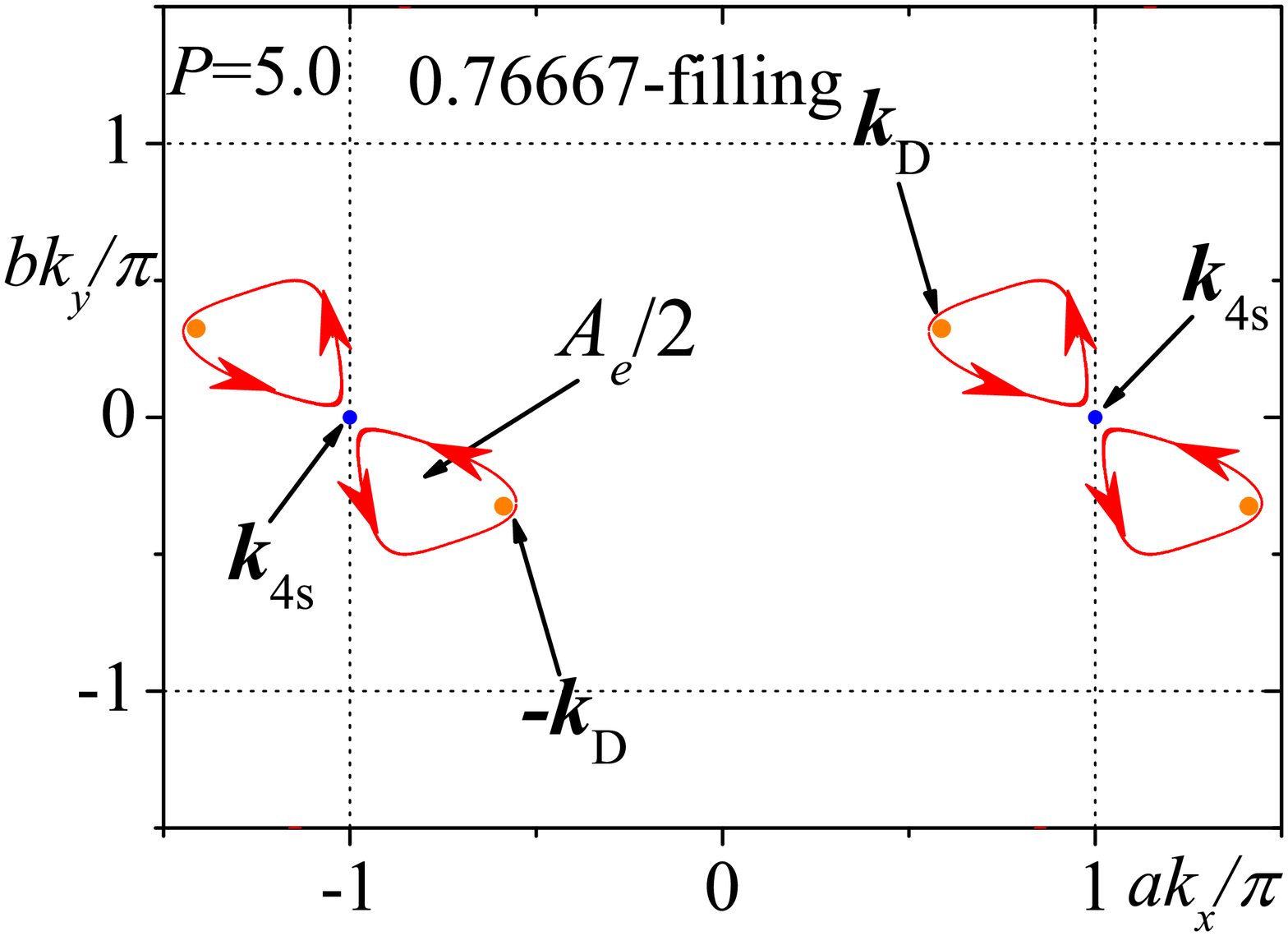}\vspace{-0.5cm}
\begin{flushleft} \hspace{0.5cm}(b) \end{flushleft}\vspace{-0.8cm} 
\includegraphics[width=0.44\textwidth]{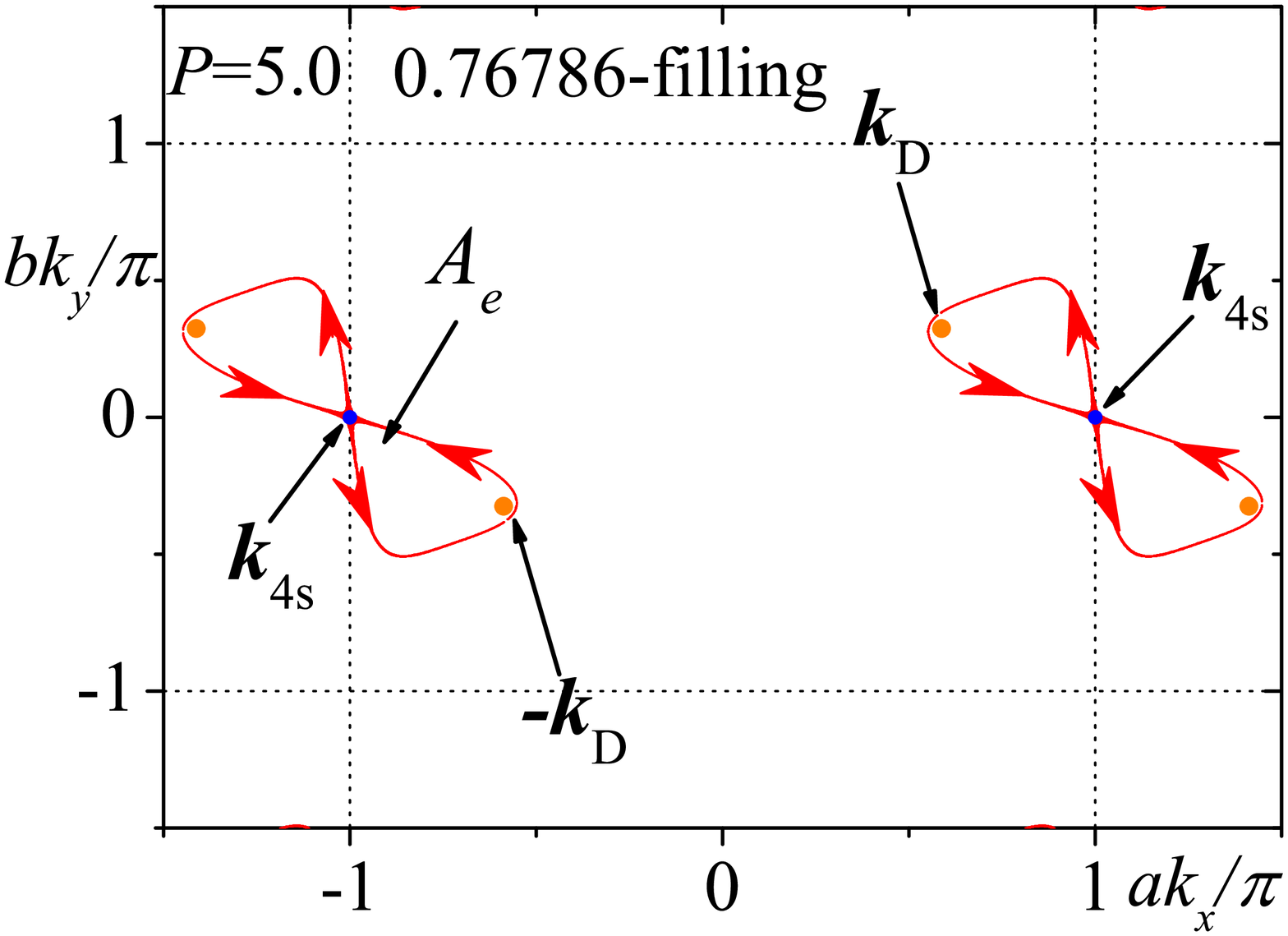}\vspace{-0.5cm}
\begin{flushleft} \hspace{0.5cm}(c) \end{flushleft}\vspace{-0.8cm}
\includegraphics[width=0.44\textwidth]{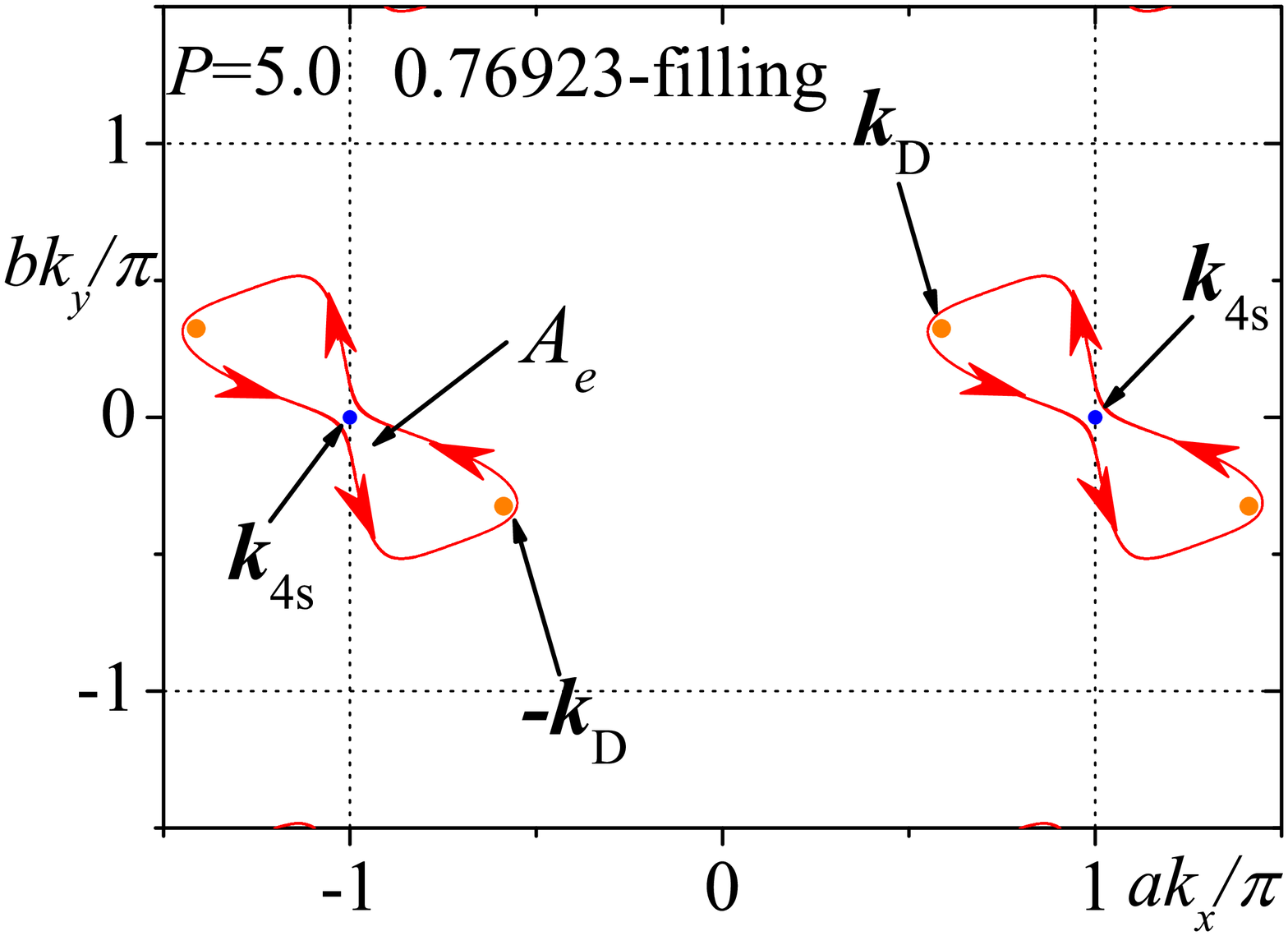}\vspace{-0.5cm}
\end{center}
\caption{ (Color online) 
Fermi surfaces at $P=5.0$ (a) 
0.76667-filling (b) 0.76786-filling and (c) 0.76923-filling in the extended 
zone, where $\pm{\bf k}_{\rm D}$ are Dirac points and 
${\bf k}_{\rm 4s}$ is the saddle point of the fourth band. Red arrows are for the direction of the orbital motion for electrons in the magnetic field in the semiclassical picture. 
We obtain $A_e/A_{\rm BZ} \simeq 0.0668$ in (a), 
$A_e/A_{\rm BZ}\simeq 0.0716$ in (b) and $A_e/A_{\rm BZ}\simeq 0.0768$ in 
(c), where $A_{\mathrm{BZ}}$ is the area of the Brillouin zone and 
$A_{{e}}$ is the sum of the area of two electron pockets [(a) and (b)] or 
the area of an electron pocket (c). 
}
\label{fig8_0}
\end{figure}

That study\cite{KH2019} has been done in two-dimensional compensated metal. 
If one hole pocket does not exist, more directly we can examine the effect of the Lifshitz transition for the dHvA oscillations. Namely, we make clear how the dHvA oscillatios are varied by the lifting 
of doubly degenerated Landau levels. 
Therefore, in this paper, we employ spinless two-dimensional tight-binding model of $\alpha$-(BEDT-TTF)$_2$I$_3$ at $P=5.0$kbar, where two electron pockets with the Dirac cones are changed to one electron pocket with a narrow neck upon increasing electron dopings (see Figs. \ref{fig2} and \ref{fig8_0}). This Fermi surface situation is realized in the doped graphene under the uniaxial strain\cite{Hasegawa2006,Rosenzweig}, in black phosphorus\cite{Kim_2}, and twisted bilayer graphene\cite{lopes,aya}. Thus, our calculation is applicable to the two-dimensional Dirac fermion systems widely.

\begin{figure}[bt]
\begin{flushleft} \hspace{0.5cm}(a)
 \end{flushleft}\vspace{-0.2cm} 
\hspace{0.1cm}\includegraphics[width=0.22\textwidth]
{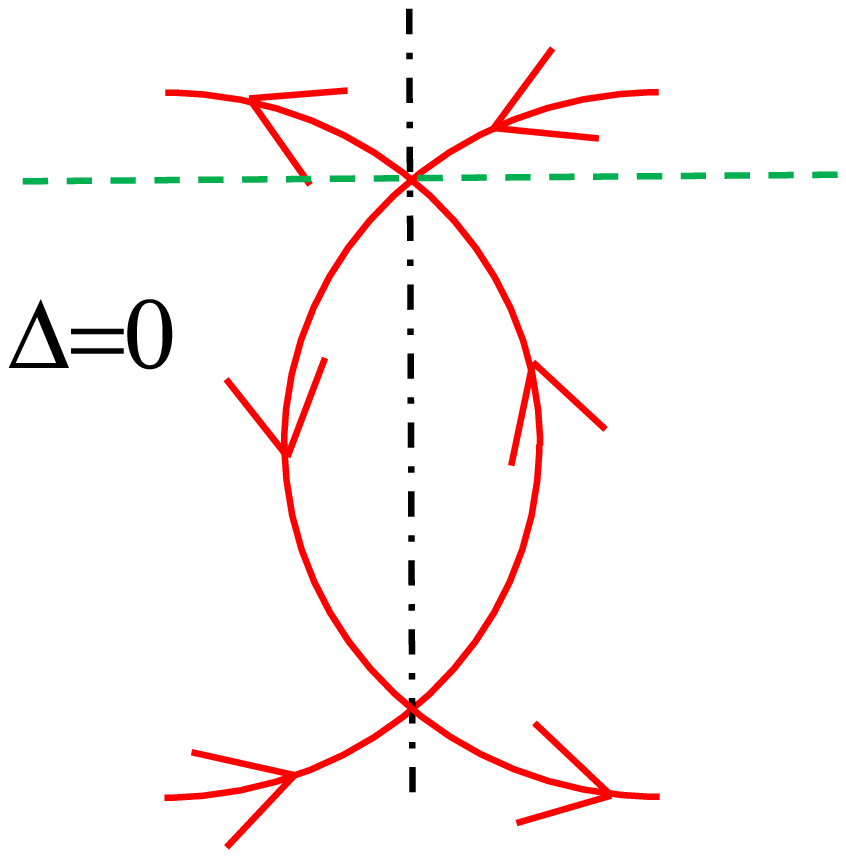}\vspace{0.4cm}
\begin{flushleft} \hspace{0.5cm}(b)
 \end{flushleft}\vspace{-0.0cm} 
\hspace{0.2cm}\includegraphics[width=0.22\textwidth]
{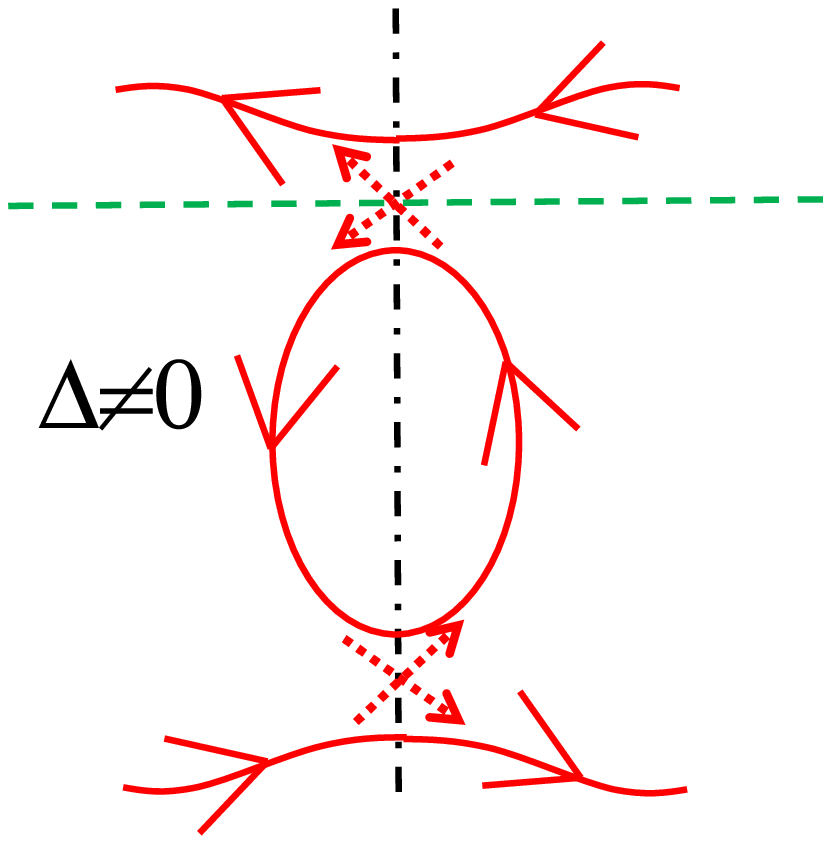}\vspace{0.4cm}
\begin{flushleft} \hspace{0.5cm}(c)
 \end{flushleft}\vspace{-0.0cm} 
\hspace{0.2cm}\includegraphics[width=0.24\textwidth]
{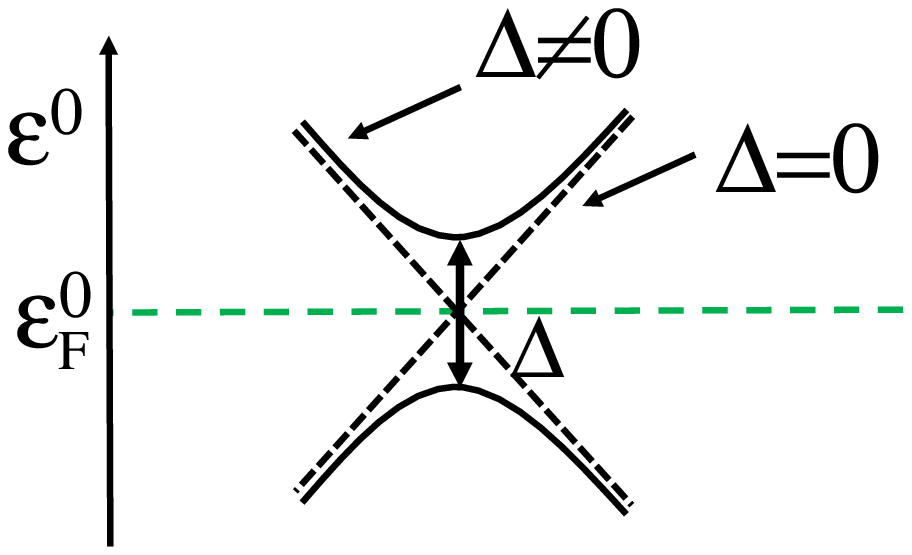}\vspace{0.4cm}
\begin{flushleft} \hspace{0.5cm}(d)
 \end{flushleft}\vspace{-0.0cm} 
\hspace{0.2cm}\includegraphics[width=0.24\textwidth]
{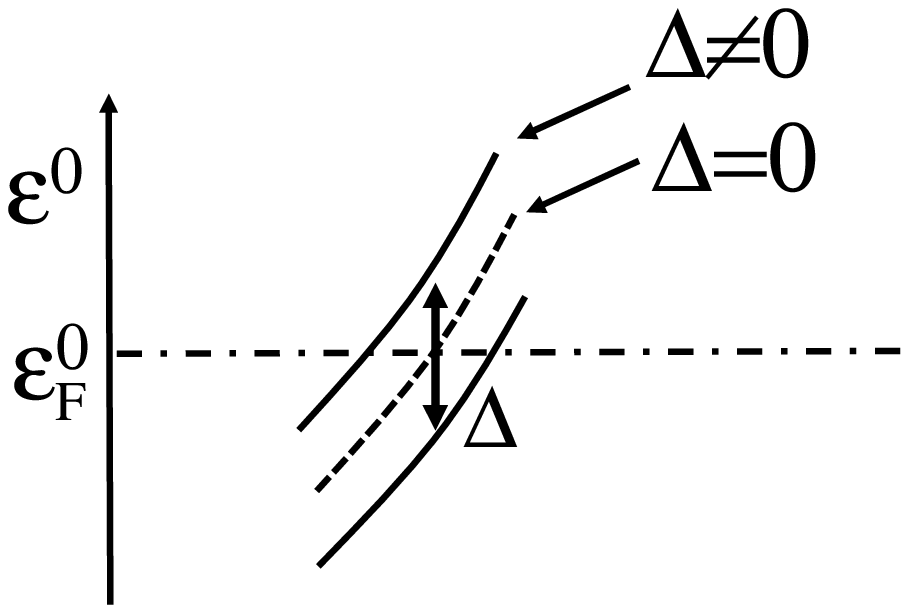}\vspace{0.4cm}
\caption{ (Color online) Schematic figures for the tunneling of electrons in the momentum space [(a) and (b)] for two-dimensional electron pocket ($\Delta=0$) and quasi-one-dimensional Fermi surface and two-dimensional electron pocket ($\Delta \neq 0$). Red lines are the Fermi surface. Red dotted arrows indicate the direction of the tunneling. Black dash-dotted lines are the Brillouin zone boundary. (c) and (d) are schematic band structures along green dash lines in (a) and black dash-dotted in (b), respectively.}\label{fig_mb}
\end{figure}

\begin{figure}[bt]
\begin{flushleft} \hspace{0.5cm}(a)
 \end{flushleft}\vspace{-0.2cm} 
\hspace{0.2cm}\includegraphics[width=0.17\textwidth]
{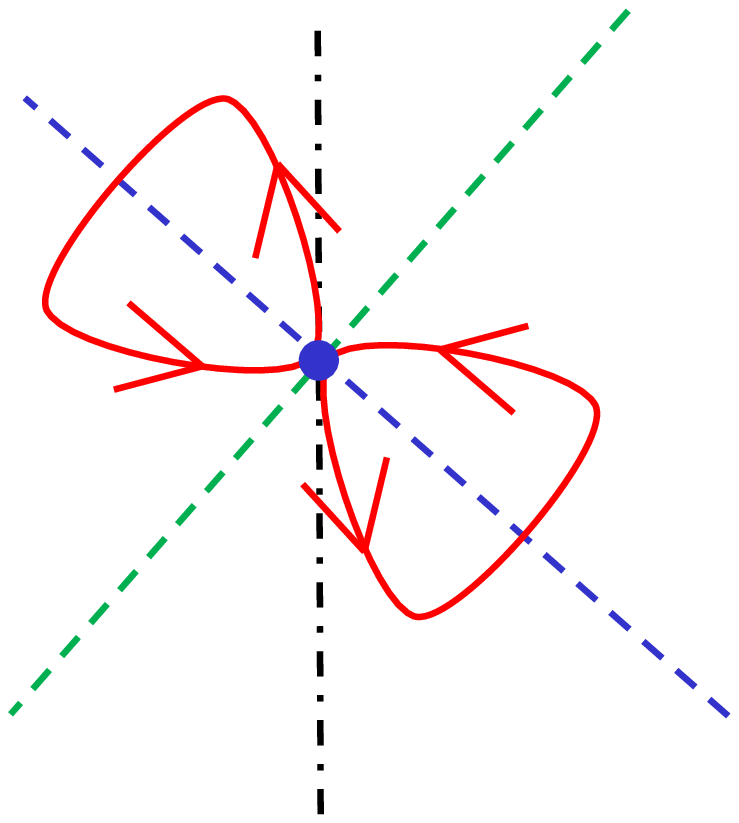}\vspace{0.4cm}
\begin{flushleft} \hspace{0.5cm}(b)
 \end{flushleft}\vspace{-0.0cm} 
\hspace{0.2cm}\includegraphics[width=0.17\textwidth]
{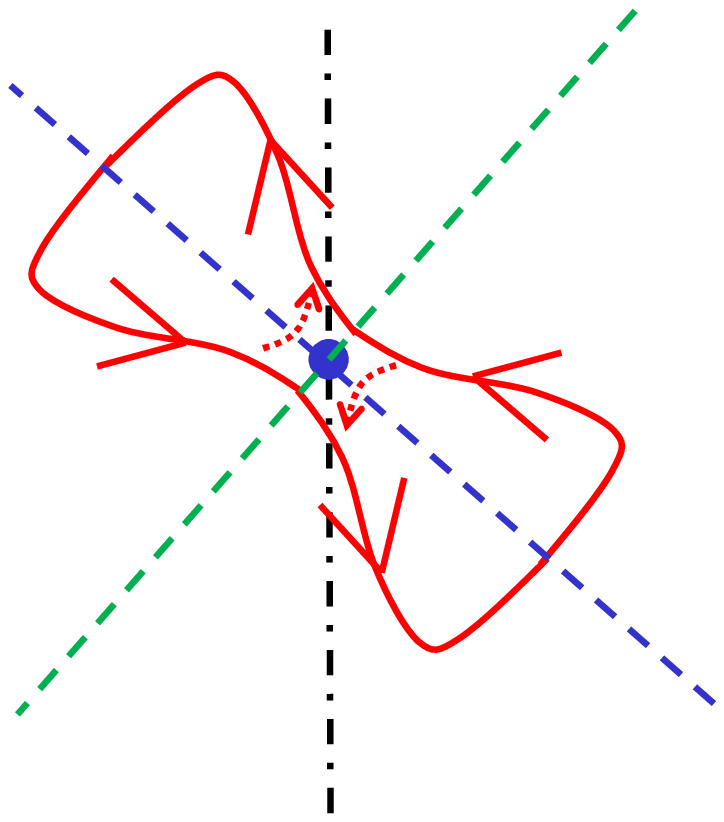}\vspace{0.4cm}
\begin{flushleft} \hspace{0.5cm}(c)
 \end{flushleft}\vspace{-0.0cm} 
\hspace{0.2cm}\includegraphics[width=0.17\textwidth]
{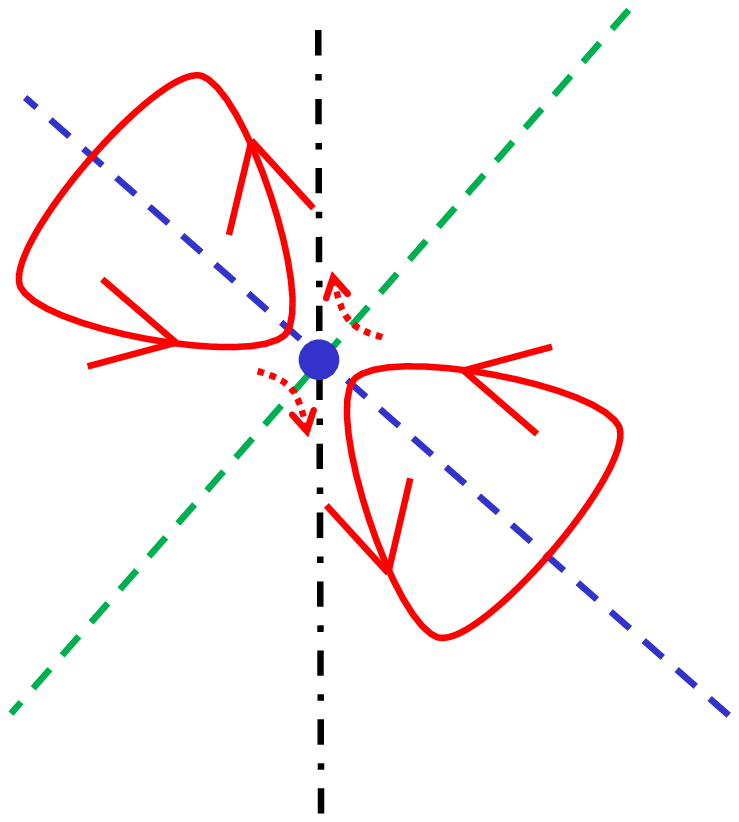}\vspace{0.4cm}
\begin{flushleft} \hspace{0.5cm}(d)
 \end{flushleft}\vspace{0.2cm} 
\hspace{-0.3cm}\includegraphics[width=0.33\textwidth]
{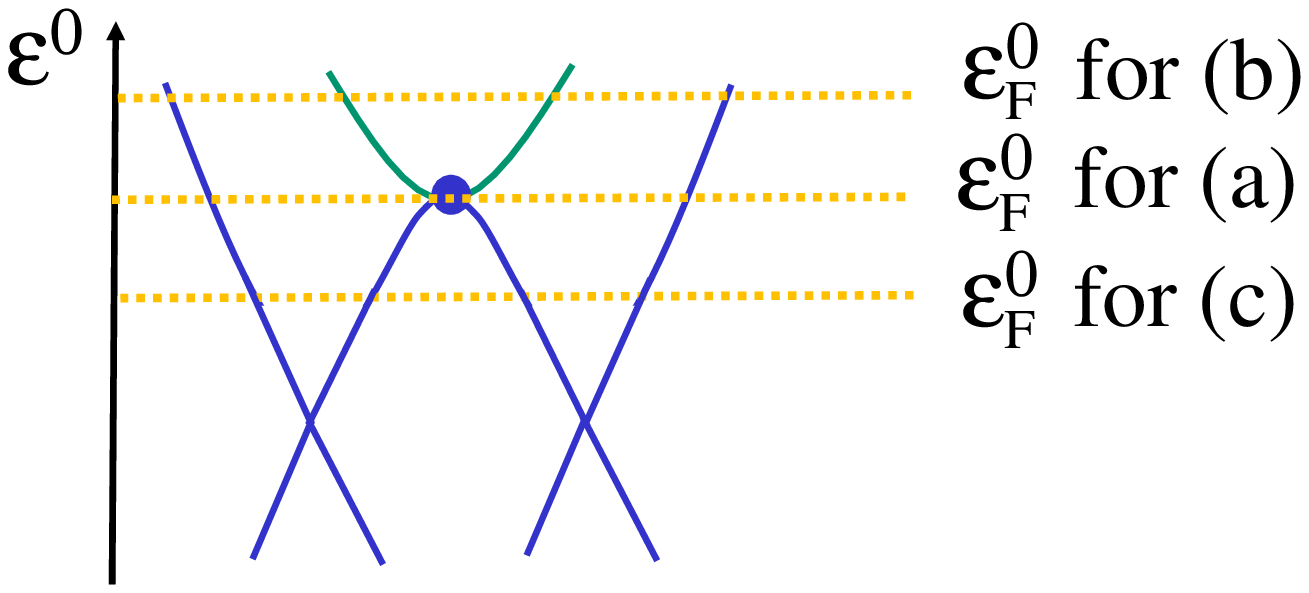}\vspace{0.6cm}
\caption{ (Color online) Schematic figures for the tunneling of electrons in the momentum space [(a), (b) and (c)] for the model used in this study. Red lines are the Fermi surface. Red dotted arrows indicate the direction of the tunneling. Black dash-dotted lines are the Brillouin zone boundary. Blue dots are the saddle points. In (d), green and blue lines are the schematic band structures along green dash and blue dash lines in (a), (b) and (c), respectively. The dispersions of green line and blue line have a minimum and a maximum at the saddle point, respectively.}\label{fig_mb2}
\end{figure}

Lifshitz and Kosevich\cite{shoenberg,LK} have derived semiclassically the standard LK formula for the dHvA oscillations, which is explained in Appendix \ref{LKformula_dHvA}, where the frequency is proportional to the extremal cross-sectional area of the Fermi surface. Since in the Fermi surface of Fig. \ref{fig8_0} the field-induced quantum tunneling is expected, a new period corresponding to that effective closed area may be additionally seen in the dHvA oscillations. This phenomena is semiclassically called the magnetic breakdown\cite{shoenberg}. The semiclassical network model\cite{Pippard62,Falicov66,blount} for the magnetic breakdown is conventionally used, in which the probability amplitude of the tunneling is introduced into the LK formula as parameters. But, since the network model\cite{Pippard62,Falicov66} has been constructed based on electron tunneling such as Fig. \ref{fig_mb}, that model may not be applicable to the Fermi surface of Figs. \ref{fig8_0} and \ref{fig_mb2}, which will be explained in the next section. Therefore, in this paper, we study the dHvA oscillations by quantum mechanical calculations by using the tight-binding model with the Peierls phase.

\begin{figure}[bt]
\begin{center}
\begin{flushleft} \hspace{0.5cm}
(a) 
\end{flushleft}\vspace{-0.3cm} 
\includegraphics[width=0.46\textwidth]{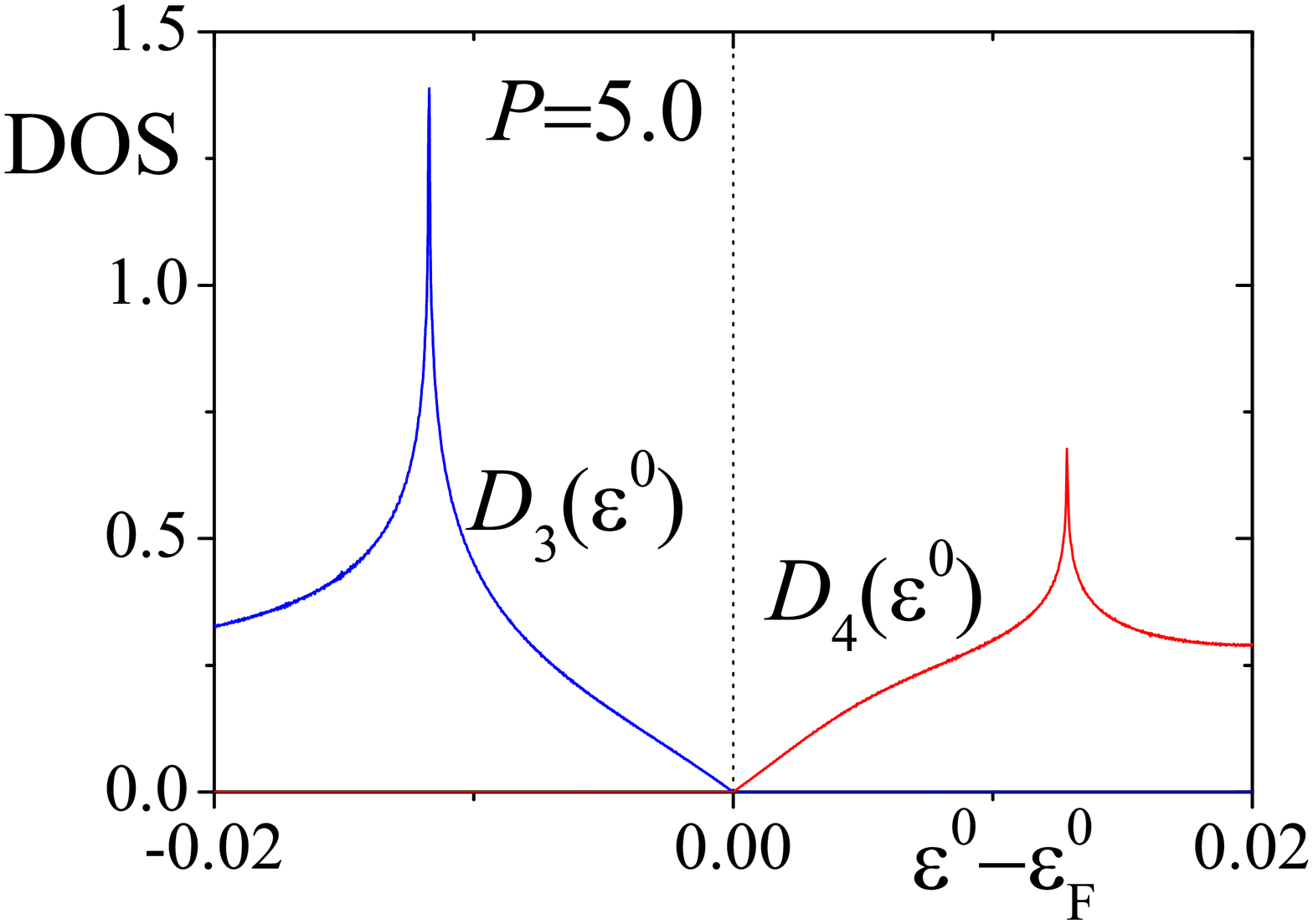}\vspace{-0.3cm}
\begin{flushleft} \hspace{0.5cm}(b) \end{flushleft}\vspace{-0.5cm} 
\includegraphics[width=0.46\textwidth]{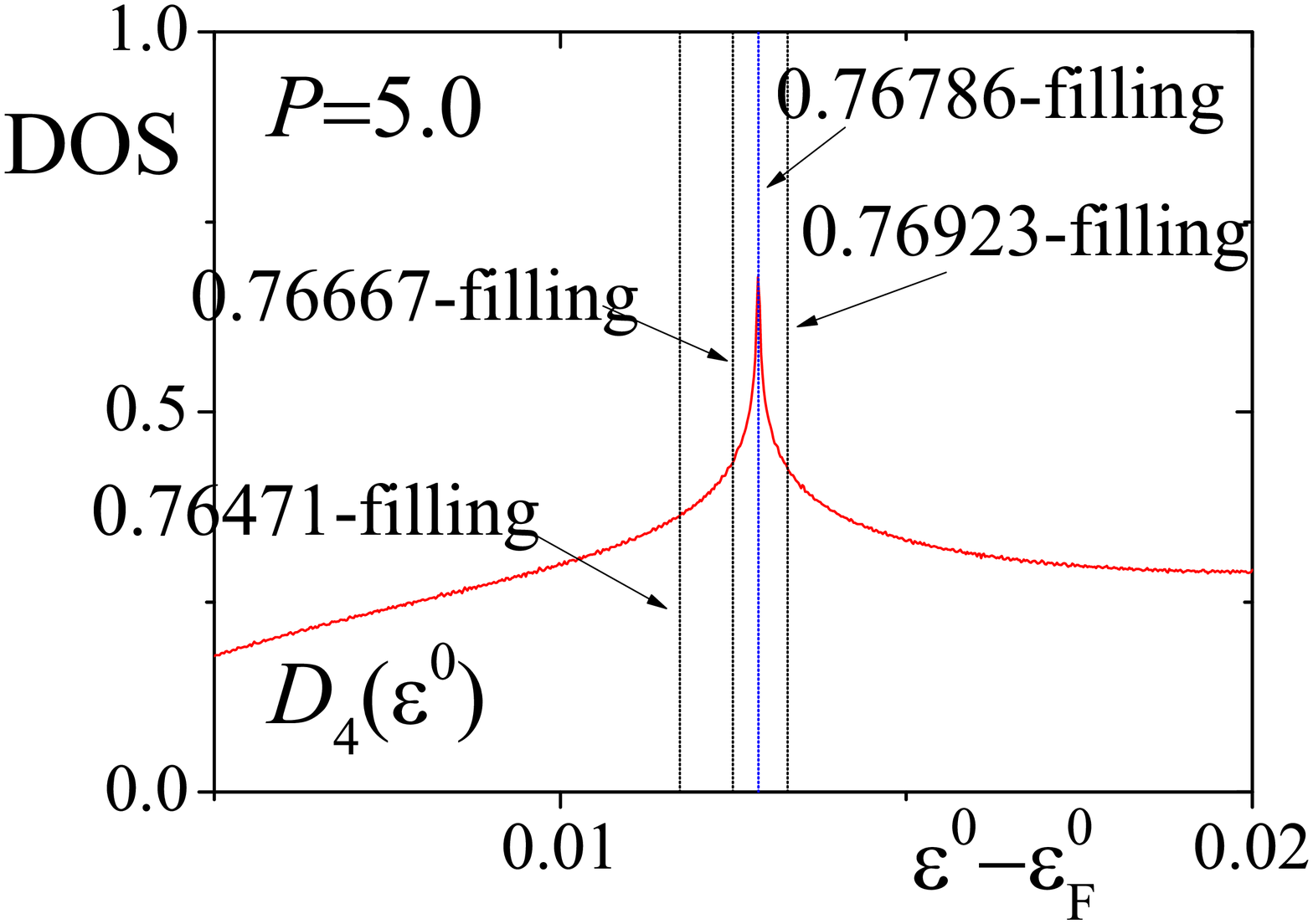}\vspace{-0.3cm}
\end{center}
\caption{
 (Color online) 
(a) The density of states [$D_3(\varepsilon ^0)$ and $D_4(\varepsilon ^0)$] of the third band and the fourth band  $H=0$ and $P=5.0$ as a function of the energy $(\varepsilon^0)$ measured from $\varepsilon^0_{{\rm F}}$. (b) is an enlarged figure of $D_4(\varepsilon ^0)$.
}
\label{fig9_N}
\end{figure}

\section{Tight-binding model for $\alpha$-(BEDT-TTF)$_2$I$_3$ and these Fermi surafaces}

$\alpha$-(BEDT-TTF)$_2$I$_3$\cite{review,review2,Kajita2014} is known as one of quasi-two-dimensional organic superconductors. The realization of the massless Dirac fermions have been theoretically shown by using the tight-binding model\cite{Katayama2006} and the first-principle band calculations\cite{Kino,Kino2009,Alemany2012}. Experimentally, massless Dirac fermions have been confirmed under the pressure. For example, the temperature ($T$) dependence in the resistivity is very small\cite{kajita1992,tajima2000}. The electronic specific heat is almost proportional to $T^2$\cite{Konoike2012}. The $N=0$ Landau level and the phase of the Landau level ($\gamma=0$) have also been shown from the interlayer magnetotransport\cite{Osada2008} and from the Shubnikov-de Hass (SdH) oscillations\cite{Tajima2013}, respectively.

In this study, we ignore the three-dimensionality of $\alpha$-(BEDT-TTF)$_2$I$_3$ due to the smallness of the interlayer coupling. 
Four energy bands are described by the highest occupied molecular orbits (HOMO) of four BEDT-TTF molecules in the tight-binding model. 
A metal-insulator transition happens at the ambient pressure and 
at low pressures. This transition is attributed to the charge ordering\cite{KF1995,Seo2000,takano2001,Woj2003}. 
However, it has been observed that the metal-insulator transition is suppressed and the metallic state is realized by the hole dopings under the pressure of 17 kbar\cite{Tajima2013} or the electron dopings under the pressure of more than 15 kbar\cite{Tiss2017}. The SdH oscillations have been also observed, where the frequencies are 1.4 T and 9.18 T in \cite{Tajima2013} and 2 T and 8.5 T in \cite{Tiss2017}. 
Since we focus on that metallic situation, we neglect the electron interactions.

In the tight-binding model employed in this paper, the interpolation formula\cite{Kobayashi2004,KH2017,KH2019} for the transfer integrals based on the extended H\"uckel method\cite{Mori1984,Kondo2005} has been used. In these band parameters, at low uniaxial pressure ($P \lesssim 3.0$kbar $=0.3$GPa) in 
$\alpha$-(BEDT-TTF)$_2$I$_3$, the energy at the Dirac points is smaller than the maximum energy of the lower band. Then the Fermi surface of the non-doping system consists of one hole pocket and one or two electron pockets, i.e., the system becomes the compensated metal. 
At $P\simeq 0.2$ kbar, the Lifshitz transition\cite{Lifshitz} occurs, where one electron pocket is transformed into two electron pockets. 
Hereafter, we take eV and kbar as the units of transfer integrals and the pressure, respectively. In this study, we fix $P=5.0$ kbar and change the electron filling. 
The third and fourth bands from the bottom are shown in Fig. \ref{fig2}. 
The Fermi surfaces for 0.76667-filling (two electron pockets), 
0.76786-filling (Lifshitz transition) and 0.76923-filling (one electron pocket) are shown in Fig. \ref{fig8_0}. 
At 0.76667-filling, there are two electron pockets with the same area, as shown in Fig. \ref{fig8_0} (a). 
When the electron filling is increased, the Lifshitz transition occurs at the 0.76786-filling, as shown in Fig. \ref{fig8_0} (b). At 0.76923-filling, 
there is one large electron pocket with a narrow point, as shown in Fig. \ref{fig8_0} (c). We numerically calculate 
the magnetizations changing the filling from the 0.76667-filling to the 0.76923-filling at $P=5.0$. Since $\alpha$-(BEDT-TTF)$_2$I$_3$ has $\frac{3}{4}$-filling, the electron dopings are needed 
to confirm our calculations experimentally. 

In the Fermi surface of Fig. \ref{fig8_0}, the saddle point at $(k_x, k_y)=(\pi/a, 0)$ or $(-\pi/a, 0)$ is shown, which is the time-reversal invariant momentum (TRIM). The densities of states are logarithmic divergences due to the saddle point, as shown in Fig.~\ref{fig9_N}. 
Recently, the dHvA oscillations near the Lifshitz transition have been calculated\cite{Itskovsky2005} by taking the effect of the van Hove singularity due to the saddle point by using the semiclassical approximation in ref. \cite{zil_jetp1958,zil_jetp1957}. They\cite{Itskovsky2005} have found that the dHvA oscillations near the Lifshitz transition is different from the LK formula. 
The system considered by them is that a two-dimensional electron pocket is changed to the quasi-one-dimensional Fermi surface where broadening of the Landau levels is caused, while two electron pockets are changed to one electron pocket in our model, as shown in Fig. \ref{fig8_0}.

The semiclassical network model has often been adopted in the case of Fig. \ref{fig_mb}\cite{Pippard62,Falicov66,blount}.
In that treatment the tunneling probability for the magnetic breakdown is assumed to be $P=\exp(-\frac{H_0}{H})$, where $H_0$ is called the breakdown field, $H_0 \propto \Delta^2$, and $\Delta$ is the magnitude of the energy gap\cite{blount}. 
When the energy gap is large limit, the electron tunneling between the energy band becomes zero, i.e., $P=0$. 
When the energy gap is zero ($\Delta=0$), we obtain $P=1$ and $Q=0$ ($Q$ is the reflecting probability), as shown in Fig. \ref{fig_mb} (a), where 
electrons can go straight. 
On the other hand, the case as in Fig. \ref{fig_mb2} is different from that in Fig. \ref{fig_mb}. In this case two Fermi surfaces are separated
not by the energy gap, $\Delta$, but by the saddle point.
When the chemical potential is the same as the energy at the saddle point, a semiclassical orbit in the momentum space collide head-on and cannot go straight at the saddle point. The network model may not be applied naively to the case of Fig. \ref{fig_mb2}.




\begin{figure}[bt]
\begin{flushleft} \hspace{0.5cm}(a) 
\end{flushleft}\vspace{-0.8cm}
\includegraphics[width=0.51\textwidth]{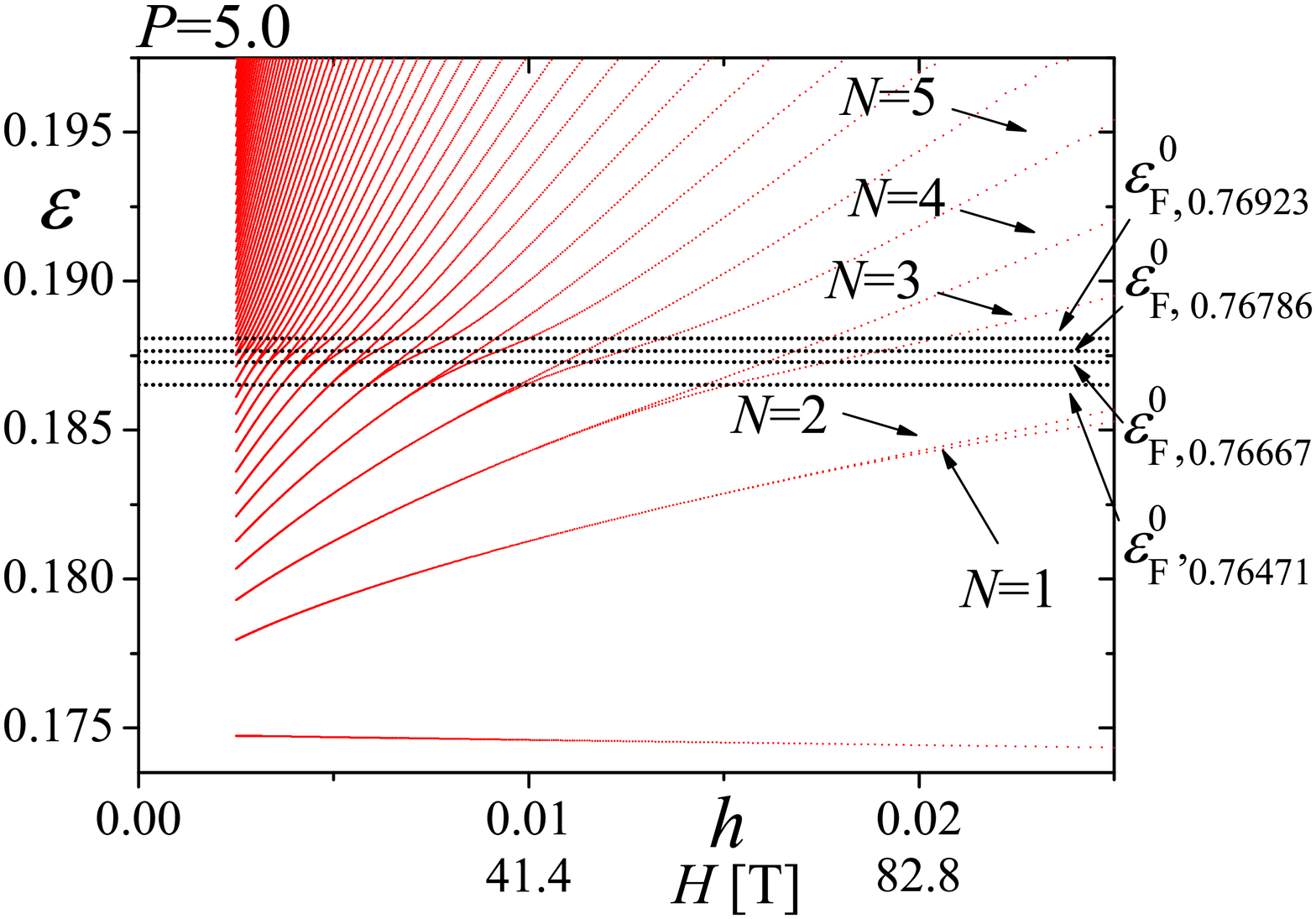}\vspace{-0.5cm}
\begin{flushleft} \hspace{0.5cm}(b) \end{flushleft}\vspace{-
0.7cm}
\includegraphics[width=0.51\textwidth]{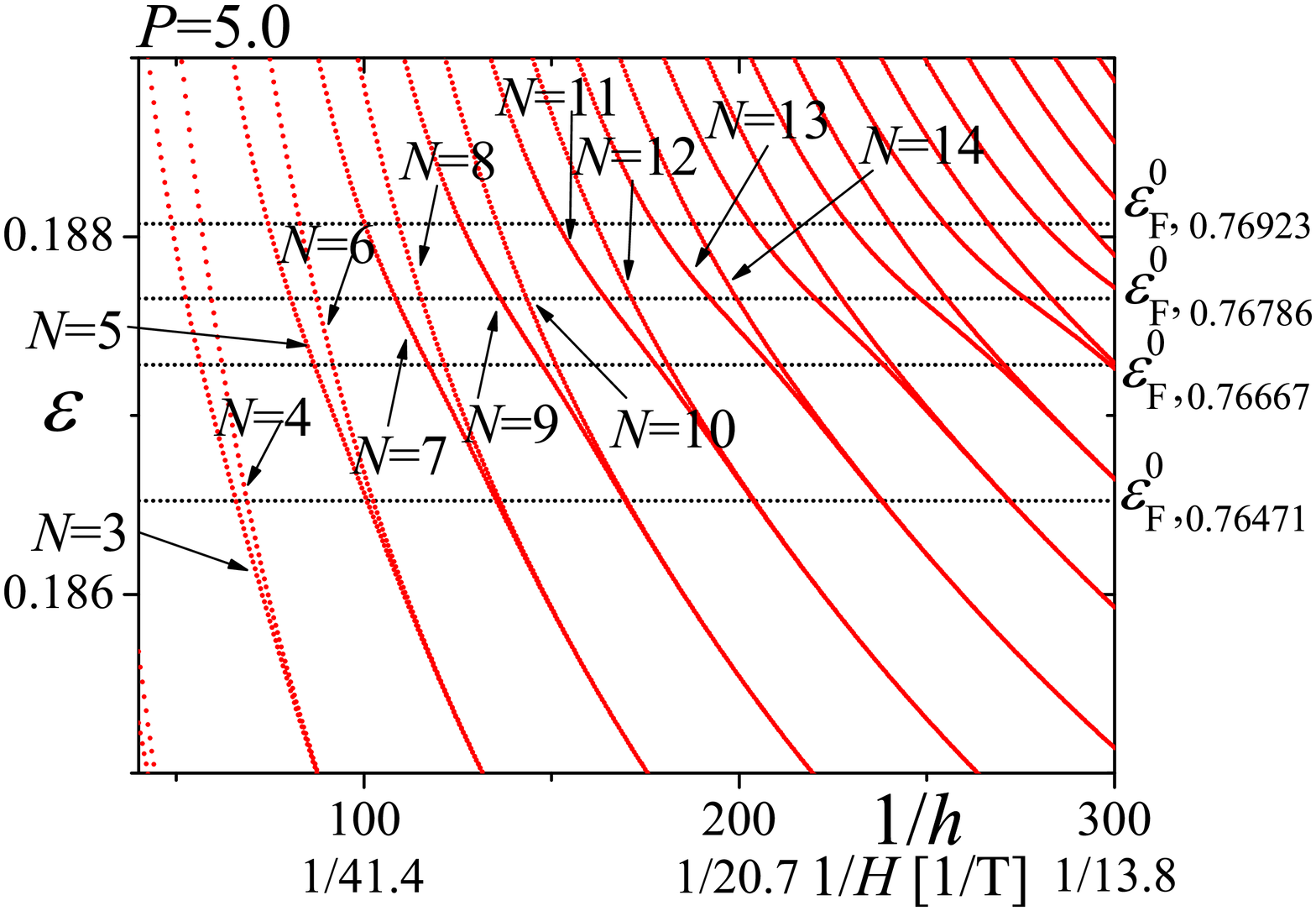}\vspace{-
0.0cm}
\caption{
(a) Energies near the Fermi energy as a function of $h$ 
at $P=5.0$, where $N$ is the index of the Landau levels for one large electron pocket. 
Black dotted lines are the Fermi energy at $h=0$ for 0.76923-filling, 0.76786-filling, 0.76667-filling and 0.76471-filling, 
respectively. (b) is an enlarged figure of (a) as a function of $1/h$. 
}
\label{fig18_0}
\end{figure}

\section{energy in the magnetic field}
When the Fermi surface is closed, the Landau levels can be obtained by using the semiclassical quantization rule\cite{Onsager}, which is explained in Appendix \ref{Appendix0}. However, in that semiclassical rule, we cannot determine the phase of the Landau levels. Moreover, that rule can not be applied to the case when the Fermi surface is not closed. 
In this paper, since we treat the case when the field-induced quantum tunneling happens, 
the energies under the magnetic field are calculated numerically in quantum mechanics. In simple cases without the tunnleing, by quantum mechanical calculations we can analytically obtain Landau levels besides phases, which is explained in Appendix \ref{Appendix01}. 

We consider the case that the uniform magnetic field is applied perpendicular to the $x-y$ plane. We neglect spins for simplicity. 
We take the ordinary Landau gauge 
\begin{equation}
{\bf A}=(Hy,0,0).
\label{ordinaryLandau}
\end{equation} 
The flux through the unit cell is given by 
\begin{eqnarray}
\Phi=abH, 
\end{eqnarray}
where $a$ and $b$ are the lattice constants. 
We use the Peierls substitution as done before\cite{KH2017}. 
We can obtain numerical solutions when the magnetic field is commensurate with the lattice period, i.e.,
\begin{eqnarray}
\frac{\Phi}{\phi_0}=\frac{p}{q}\equiv h, 
\label{Phi}
\end{eqnarray}
where $\phi_0=2\pi\hbar c/e\simeq 4.14\times 10^{-15}$ Tm$^2$ is a flux quantum, $e$ is the absolute value of the electron charge ($e > 0$), 
$c$ is the speed of light, $\hbar$ is the Planck constant divided by $2\pi$, and $p$ and $q$ are integers. Hereafter, we represent the strength of the magnetic field by $h$ of Eq. (\ref{Phi}). Since $a\simeq 9.211$~\AA \ and 
$b\simeq 10.85$\AA \ in $\alpha$-(BEDT-TTF)$_2$I$_3$\cite{review}, $h=1$ corresponds to $H\simeq 4.14\times 10^{3}$~T.

We set that the temperature is zero ($T=0$) in this study, because we consider the case when the energy level broadening due to the temperature is much smaller than the spacing of the Landau levels.

We show the energies as a function of $h$ and $1/h$, as shown in Figs. \ref{fig18_0} (a) and (b), respectively. We choose $p=2$ and $80\leq q 
\leq 600$ ($q=80, 81, \cdots, 599, 600$). The magnetic Brillouin zone are $-\frac{\pi}{a}\leq k_x<\frac{\pi}{a}$ and $-\frac{\pi}{qb}\leq k_y<\frac{\pi}{qb}$. We have checked that if $q$ is large ($q\geq 80$) which is taken in this study, the wave-number dependence of the eigenvalues $\varepsilon({i,\mathbf{k}})$ is very small. Therefore, we can safely ignore the wave-number dependence.

In Fig. \ref{fig18_0}, two Landau levels with index, $N$, are almost degenerated at the low energies or small magnetic field. When the energies or the magnetic field become larger, almost degenerated Landau levels with index $N$ are separated. The smooth separation of these Landau levels is seen when the magnetic field increase or the energy barrier between two electron pockets decreases. This separation is due to the field-induced quantum tunneling.


\section{Quantum oscillations of magnetizations}
\label{results}

The LK formula is justified when the $H$-dependence of the chemical potential can be ignored, for example, due to the electron reservoirs, the three-dimensionality, impurity effect or the thermal broadening. However, since the $H$-dependence of the chemical potential becomes large in the two-dimensional and quasi-two-dimensional systems at low temperatures, 
for these systems we have to calculate the dHvA oscillations under the conditions of the fixed electron 
number\cite{shoenberg,nakano,machida,kishigi,sandu,fortin1998,alex1996,alex2001,mineev,champel2002,KH,its2003,gvoz2003}. The dHvA oscillations in the case of the fixed electron number is different from those in the case of the fixed chemical potential. For example, 
in two-dimensional free electrons with one electron pocket or one hole pocket, the saw-tooth pattern of the dHvA oscillations in the case of the fixed electron number is inverted from that in the case of the fixed chemical potential.

\begin{figure}[bt]
\vspace{0.5cm}
\begin{flushleft} \hspace{0.5cm}(a) \end{flushleft}\vspace{-0.4cm}
\includegraphics[width=0.48\textwidth]{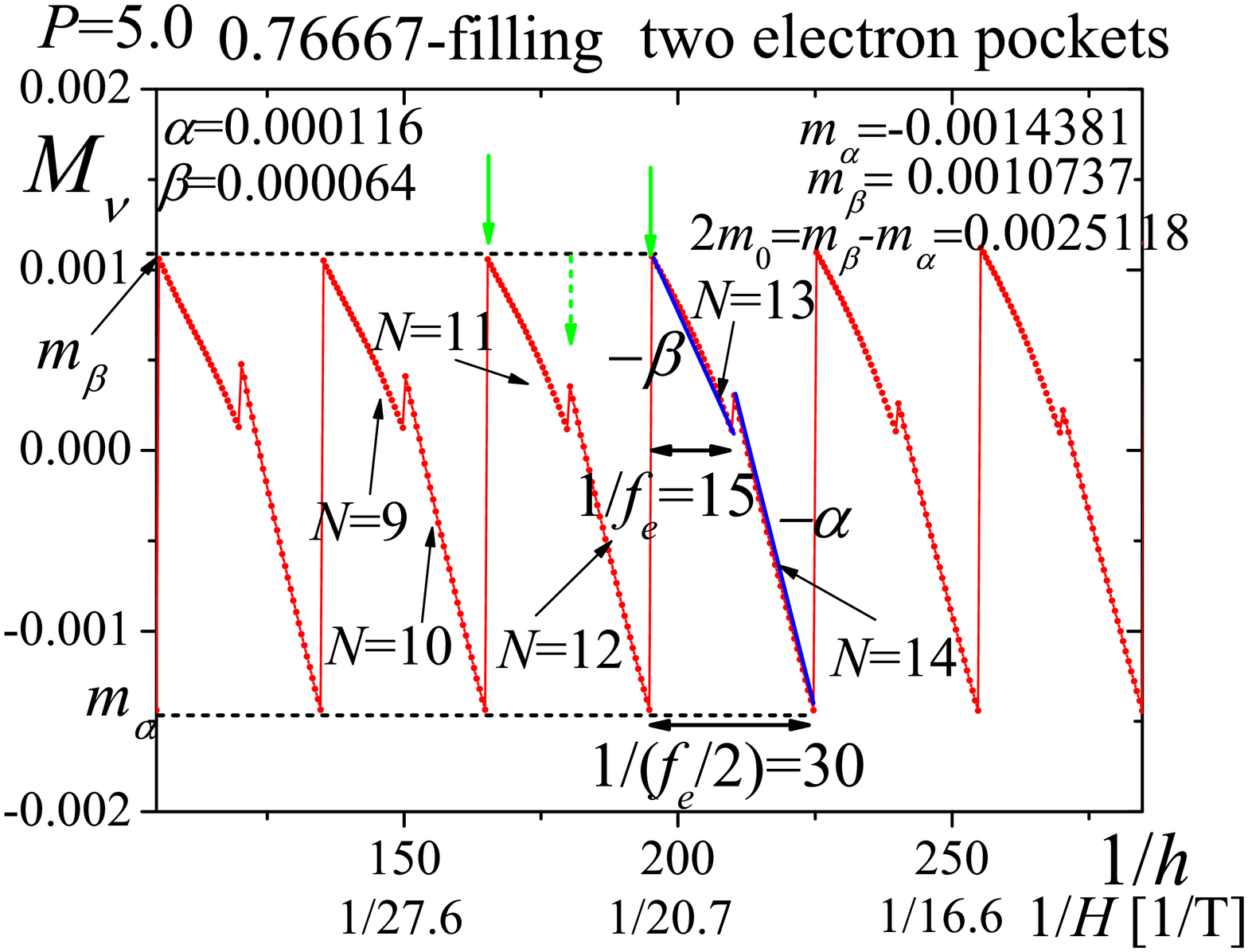}\vspace{-0.0cm}
\begin{flushleft} \hspace{0.5cm}(b) \end{flushleft}\vspace{-0.4cm}
\includegraphics[width=0.48\textwidth]{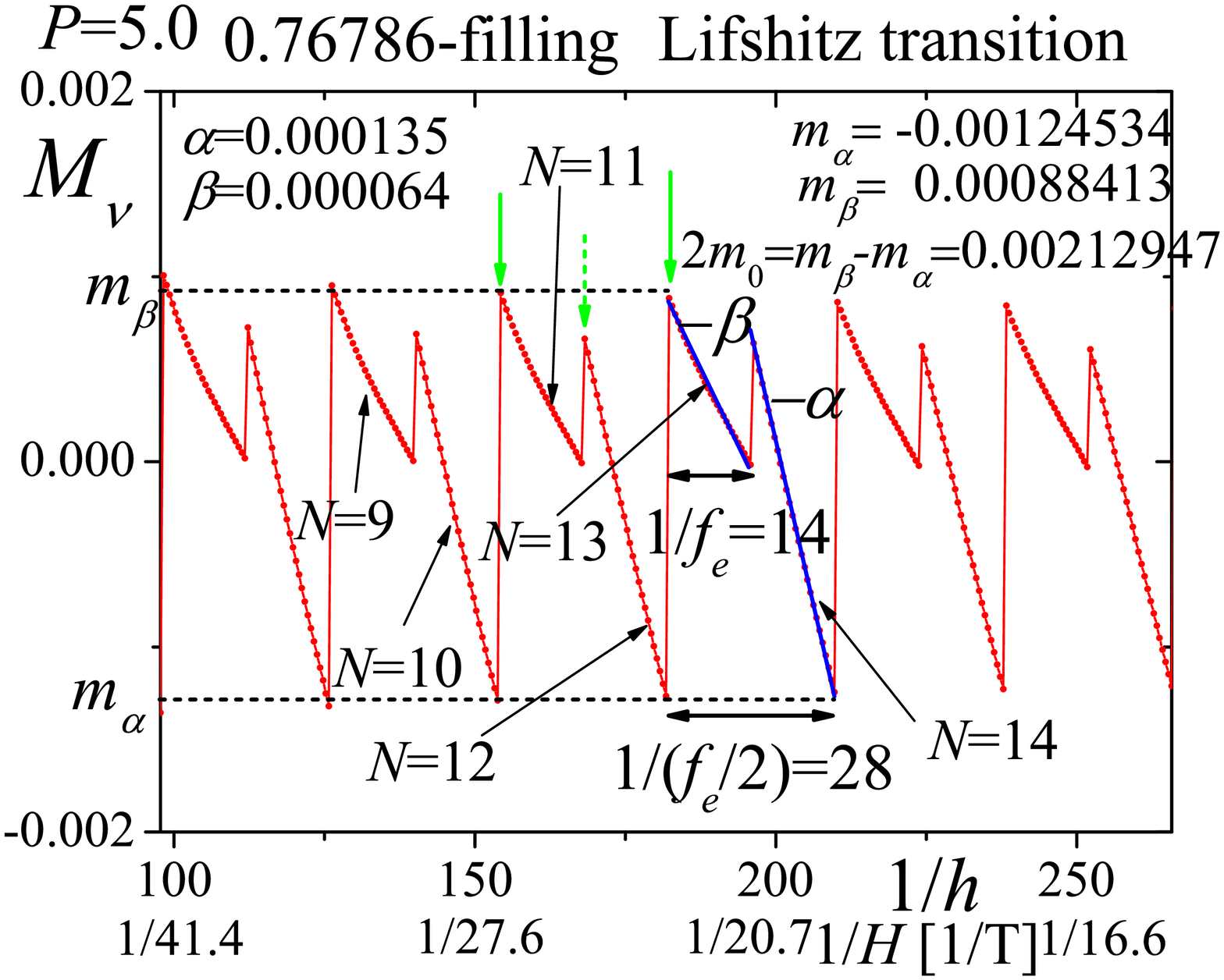}\vspace{-0.0cm}
\begin{flushleft} \hspace{0.5cm}(c) \end{flushleft}\vspace{-0.4cm}
\includegraphics[width=0.48\textwidth]{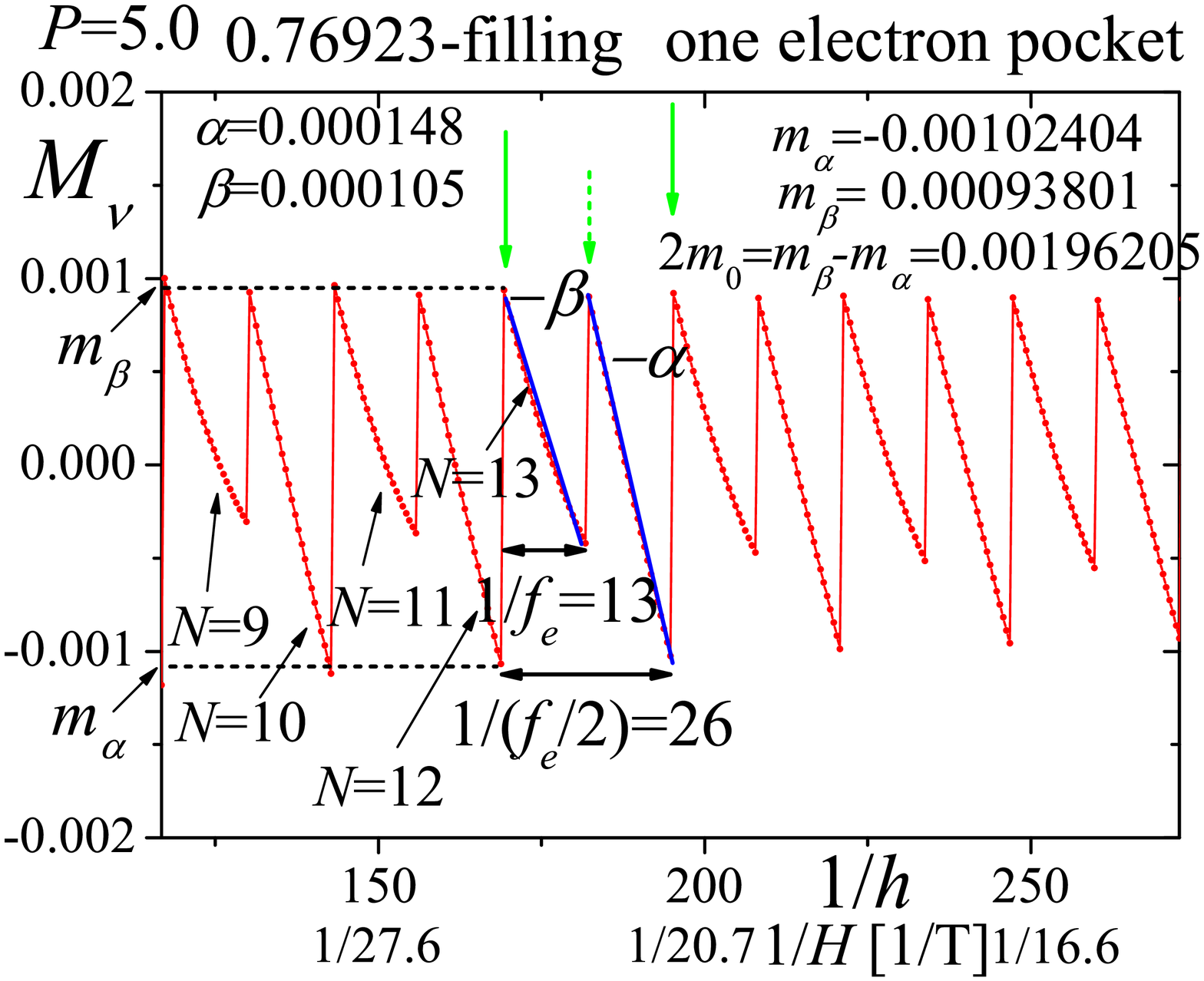}\vspace{-0.2cm}
\caption{
Magnetizations as a function of $1/h$ with fixed electron filling $\nu$ at 0.76667-filling (a), at 0.76786-filling (b), and at 0.76923-filling (c). 
}
\label{fig34}
\end{figure}

\begin{figure}[bt]
\vspace{0.5cm}
\begin{flushleft} \hspace{0.5cm}(a) \end{flushleft}\vspace{-0.4cm}
\includegraphics[width=0.5\textwidth]{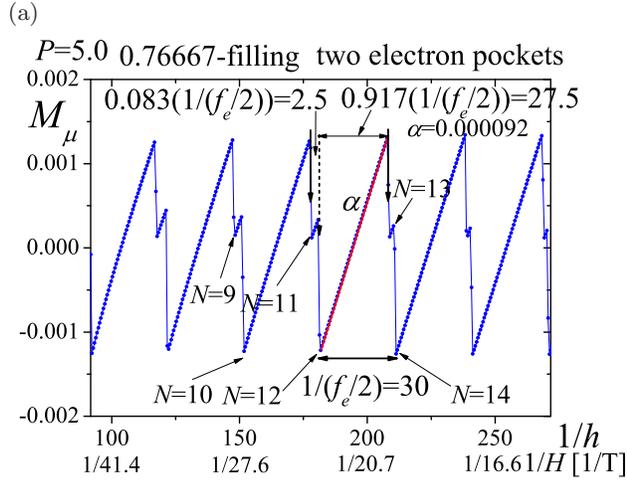}\vspace{0.3cm}
\begin{flushleft} \hspace{0.5cm}(b) \end{flushleft}\vspace{-0.4cm}
\includegraphics[width=0.5\textwidth]{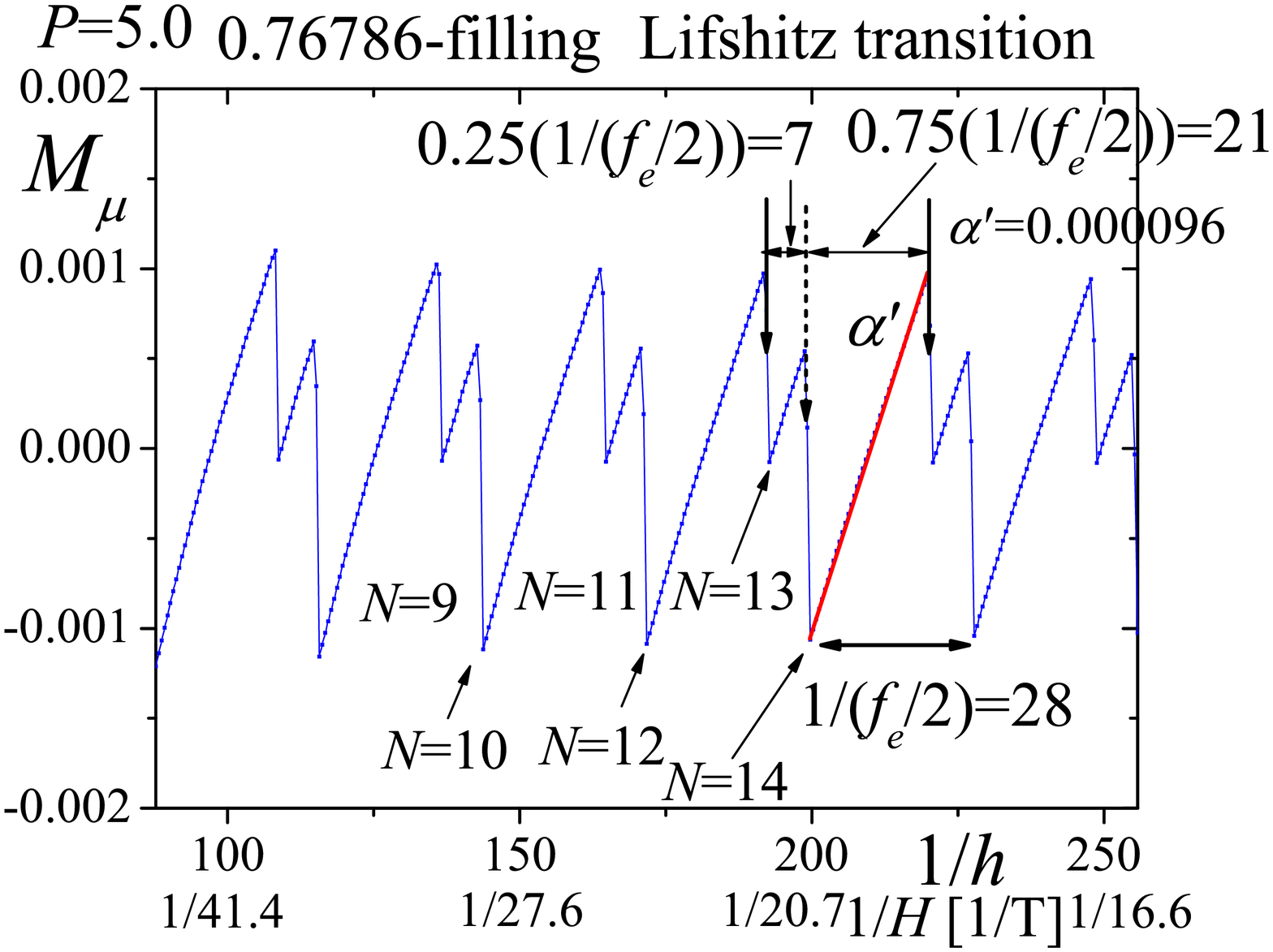}\vspace{0.3cm}
\begin{flushleft} \hspace{0.5cm}(c) \end{flushleft}\vspace{-0.4cm}
\includegraphics[width=0.5\textwidth]{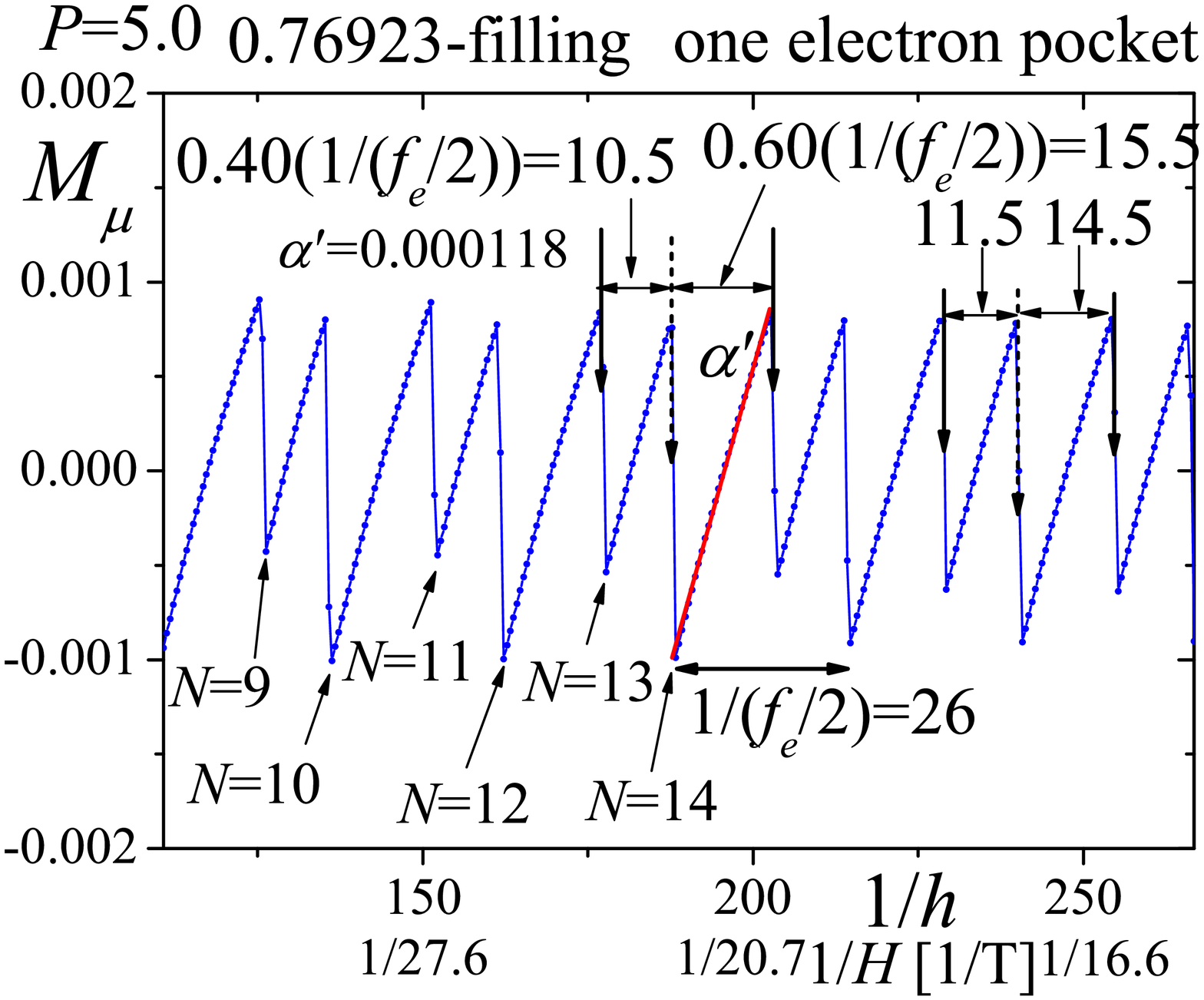}\vspace{0.3cm}
\caption{
Magnetizations as a function of $1/h$ with fixed chemical potential $\mu$ at 0.76667-filling (a), at 0.76786-filling (b), and at 0.76923-filling (c). 
}
\label{fig35}
\end{figure}

\begin{figure}[bt]
\vspace{0.5cm}
\begin{flushleft} \hspace{0.5cm}(a) \end{flushleft}\vspace{-0.6cm}
\includegraphics[width=0.5\textwidth]{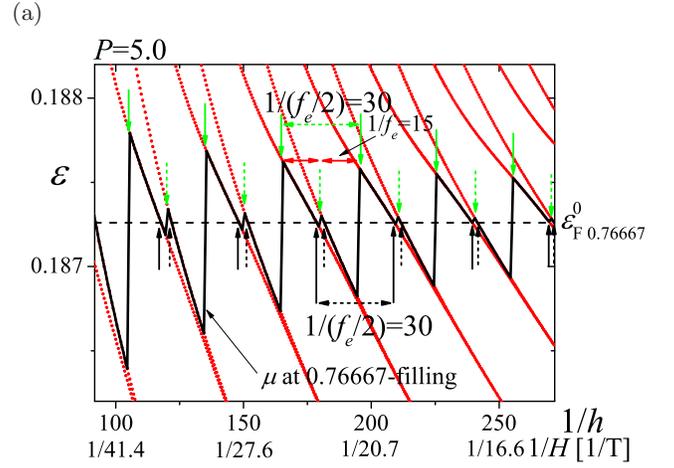}\vspace{0.3cm}
\begin{flushleft} \hspace{0.5cm}(b) \end{flushleft}\vspace{-0.6cm}
\includegraphics[width=0.5\textwidth]{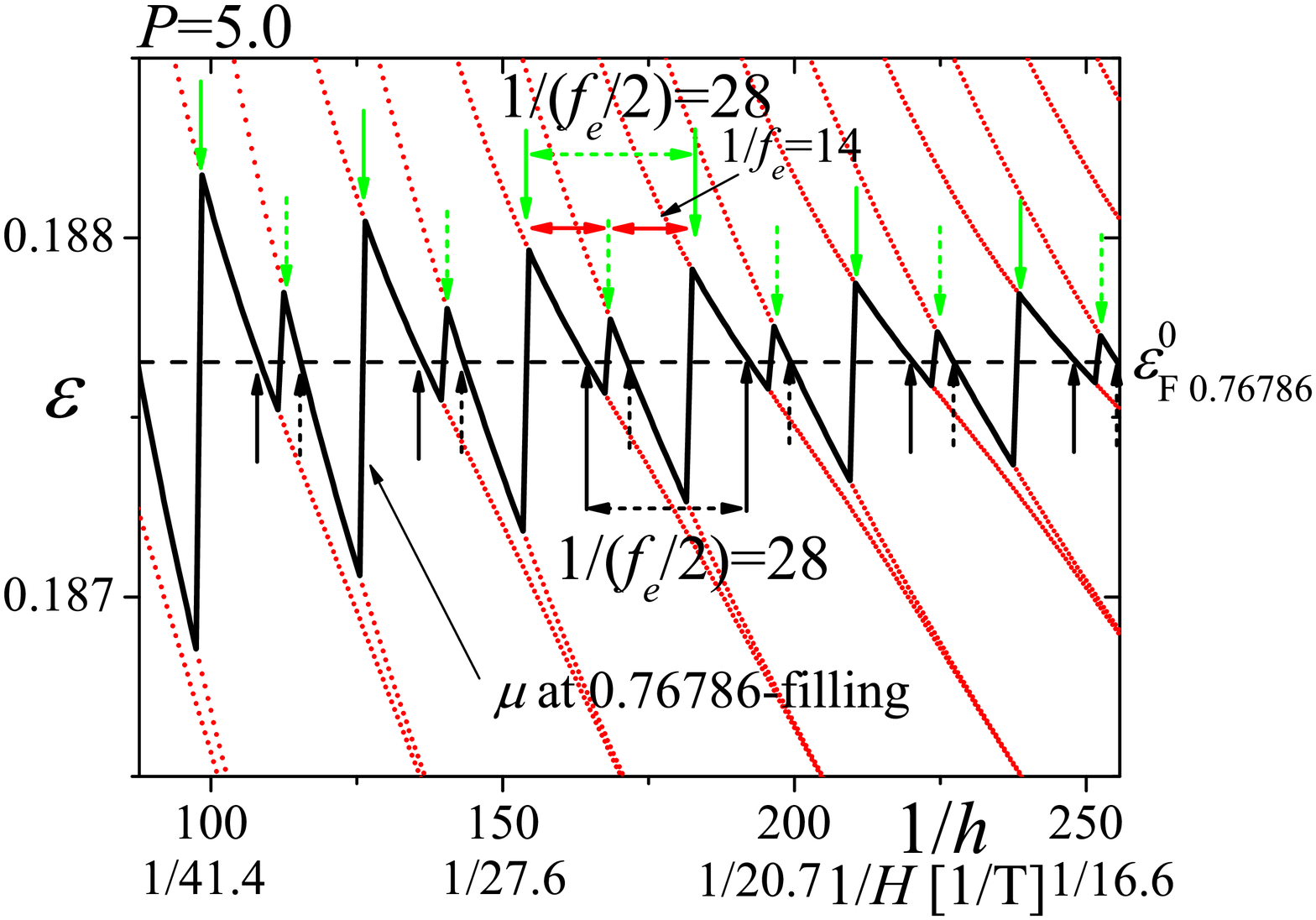}\vspace{0.3cm}
\begin{flushleft} \hspace{0.5cm}(c) \end{flushleft}\vspace{-0.5cm}
\includegraphics[width=0.5\textwidth]{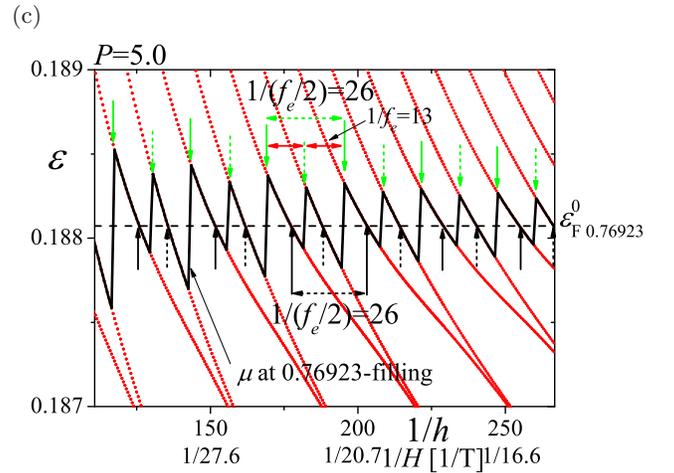}\vspace{0.0cm}
\caption{
The chemical potential $\mu$ as a function of $1/h$ at 0.76667-filling (a), at 0.76786-filling (b), and at 0.76923-filling (c) are shown by black lines.  Black dash lines are $\varepsilon_{{\rm F}, 0.76667}^0$ at 0.76667-filling (a), $\varepsilon_{{\rm F}, 0.76786}^0$ at 0.76786-filling (b), and $\varepsilon_{{\rm F}, 0.76923}^0$ at 0.76923-filling (c). Red dots are the energies. In $\mu$, the jumps of the fundamental period and the additional center jump are indicated by vertical green arrows and vertical green dotted arrows, respectively. In $\varepsilon_{{\rm F}}^0$ and the energies, the crossings for the fundamental and the additional periods are depicted by vertical black arrows and vertical black dotted arrows, respectively. 
}
\label{fig36}
\end{figure}

We calculate the magnetizations ($M_{\nu}$ and $M_{\mu}$) 
from the total energies ($E_{\nu}$ and $E_{\mu}$) in two situations; the fixed electron filling, $\nu$, and fixed chemical potential, $\mu$, respectively. The method of the calculations is explained in appendix \ref{appendix_m}. When dopings are induced by substitution, the electron filling is fixed, whereas the chemical potential is fixed when dopings are induced by electric field. We show $M_{\nu}$, $M_{\mu}$ and energies as a function of $1/h$ in the systems with two electron pockets (0.76667-filling), at Lifshitz transition (0.76786-filling), and with one electron pocket (0.76923-filling) in Figs \ref{fig34}, \ref{fig35}, and \ref{fig36}, respectively.


In Figs. \ref{fig34} (a), (b), and (c), the jumps of the fundamental period of $2/f_e$ and the additional center jump are seen in $M_{\nu}$, which are caused by the jumps of $\mu$, as shown in Figs. \ref{fig36} (a), (b), and (c). The magnetizations and $\mu$ jump at the same time. 
In Fig. \ref{fig34} (a), the center jump of $M_{\nu}$ (vertical green dotted arrows) is small. This small jump is caused by the small center jump of $\mu$ (vertical green dotted arrows) in Fig. \ref{fig36} (a) which comes from the small lifting of degenerated two Landau levels. 
Then, two electron pockets exist near each other [Fig. \ref{fig8_0} (a)]. 
In Fig. \ref{fig34} (b) [at the Lifshitz transition, where the Fermi surface is shown in Fig. \ref{fig8_0} (b)], the center jump of $M_{\nu}$ (vertical green dotted arrows) becomes larger, because the lifting of doubly degenerated Landau levels is larger and the center jump of $\mu$ is larger, as shown in Fig. \ref{fig36} (b). In Fig. \ref{fig34} (c), the center jump of $M_{\nu}$ becomes larger than that of Fig. \ref{fig34} (b), where there is one electron pocket with the narrow neck, as shown in Fig. \ref{fig8_0} (c).  The reason why the additional jump always happens at center is that the degeneracy of the Landau levels is proportional to the strength of the magnetic field. If the electron filling will be increased more than 0.76923, the jumps of the half period of $1/f_e$ will only appear.


Next, we discuss $M_{\mu}$ where $\mu$ is fixed as a function of $h$. Thus, 
$\mu=\varepsilon_{{\rm F},0.76667}^0$, $\varepsilon_{{\rm F},0.76786}^0$, and $\varepsilon_{{\rm F},0.76923}^0$, where $\varepsilon_{{\rm F},0.76667}^0$, $\varepsilon_{{\rm F},0.76786}^0$, and $\varepsilon_{{\rm F},0.76923}^0$ are the Fermi energies at 0.76667-filling, 0.76786-filling, and 0.76923-filling and $h=0$. The jumps in $M_{\mu}$ [Figs. \ref{fig35} (a), (b), and (c)] are caused by the crossings of the Landau levels and the fixed chemical potentials [Figs. \ref{fig36} (a), (b), and (c)]. 
In Fig. \ref{fig35} (a), a jump in $M_{\mu}$ is separated into two jumps with two kind of periods (2.5 and 27.5), where the degenerated two Landau levels are lifted a little, as shown in Fig. \ref{fig36} (a), and two electron pockets exist near each other, as shown in Fig. \ref{fig8_0} (a). In Fig. \ref{fig35} (b) (at the Lifshitz transition), the spacing between separated jumps in $M_{\mu}$ becomes large, where there are two jumps with periods of 7 and 21. In Fig. \ref{fig35} (c), the spacing becomes large, where 
two periods are 10.5 and 15.5, respectively. Namely, if the electron filling will be increased more, two periods will be the same. 


We summarize the results obtained in $M_{\nu}$ and $M_{\mu}$ upon changing the filling. When there exist two electron pockets at a far distance with large tunneling barrier, the wave forms of the dHvA oscillations are simple saw-tooth with the period of $2/f_e$ and these in $M_{\nu}$ and $M_{\mu}$ are inverted each other. Upon increasing the electron filling, two electron pockets become closer, and the lifting of doubly degenerated Landau levels occurs. The small center jump appears in $M_{\nu}$. One jump in $M_{\mu}$ is separated into a pair of jumps. The center jump in $M_{\nu}$ and the spacing of the separated jump in $M_{\mu}$ become larger as the electron filling increases. 
When two electron pockets meet at a saddle point (at the Lifhsitz transition), the center jump in $M_{\nu}$ and the spacing between separated jumps in $M_{\mu}$ do not change as a function of $h$, as shown in Figs. \ref{fig34} (b) and  \ref{fig35} (b), because the magnitude of tunneling barrier is almost zero. By increasing the electron filling more, the spacing of the Landau levels as a function of $1/h$ will be almost constant. Then, the wave forms of the dHvA oscillations will be almost simple saw-tooth with the period of $1/f_e$ and the neck of one electron pocket will be not narrow. Although the topology of the Fermi surface is changed [Figs. \ref{fig8_0} (a), (b), and (c)] and the density of states is divergent at the Lifshitz transition (van Hove singularity), the wave forms of the dHvA oscillations of $M_{\nu}$ and $M_{\mu}$ are varied continuously [Figs. \ref{fig34} and  \ref{fig35}]. The Lifshitz transition is seen as a crossover in the dHvA oscillations.

\begin{figure}[bt]
\vspace{0.5cm}
\begin{flushleft} \hspace{0.5cm}(a) \end{flushleft}\vspace{-0.4cm}
\includegraphics[width=0.46\textwidth]{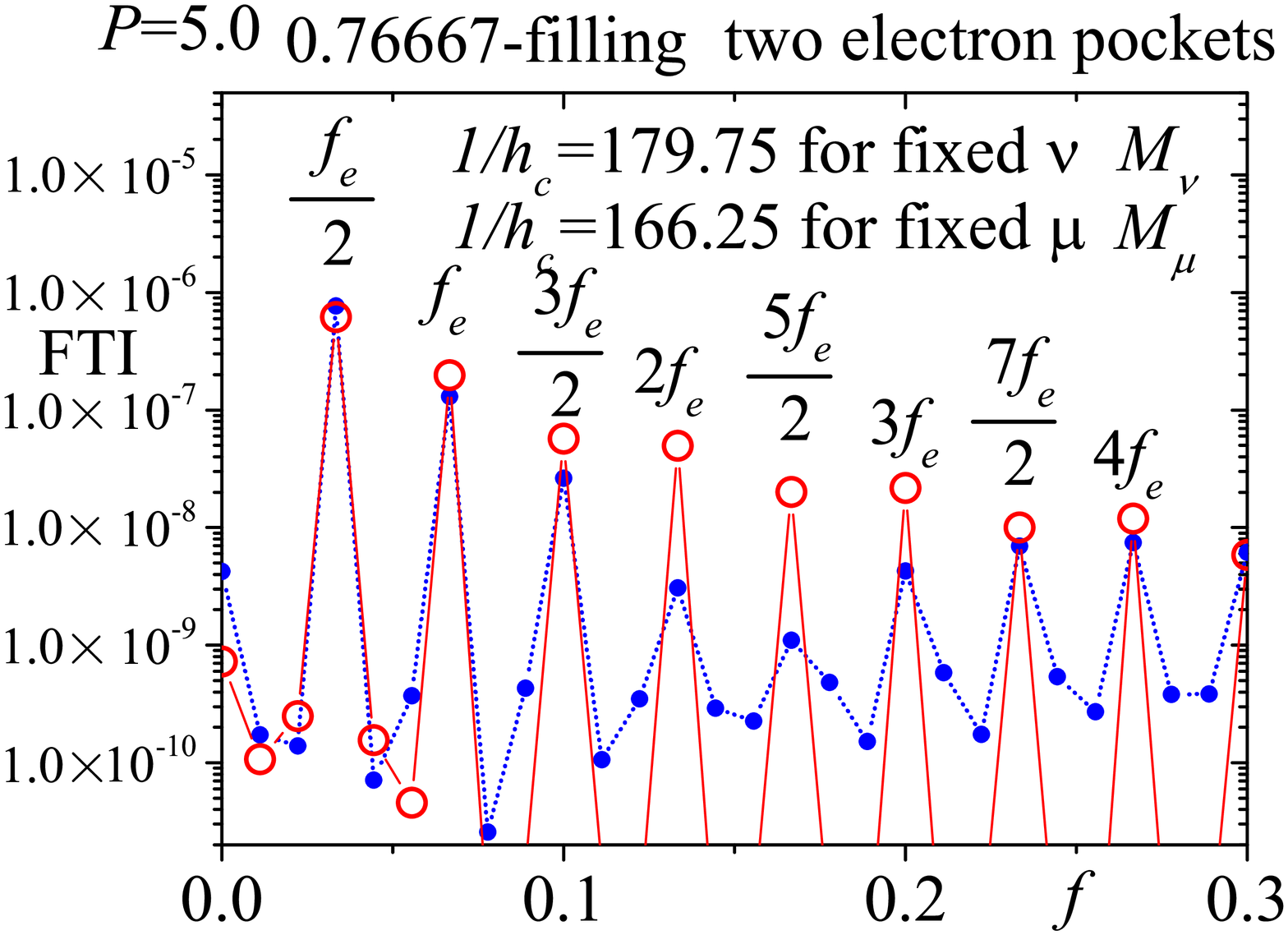}\vspace{-0.0cm}
\begin{flushleft} \hspace{0.5cm}(b) \end{flushleft}\vspace{-0.4cm}
\includegraphics[width=0.46\textwidth]{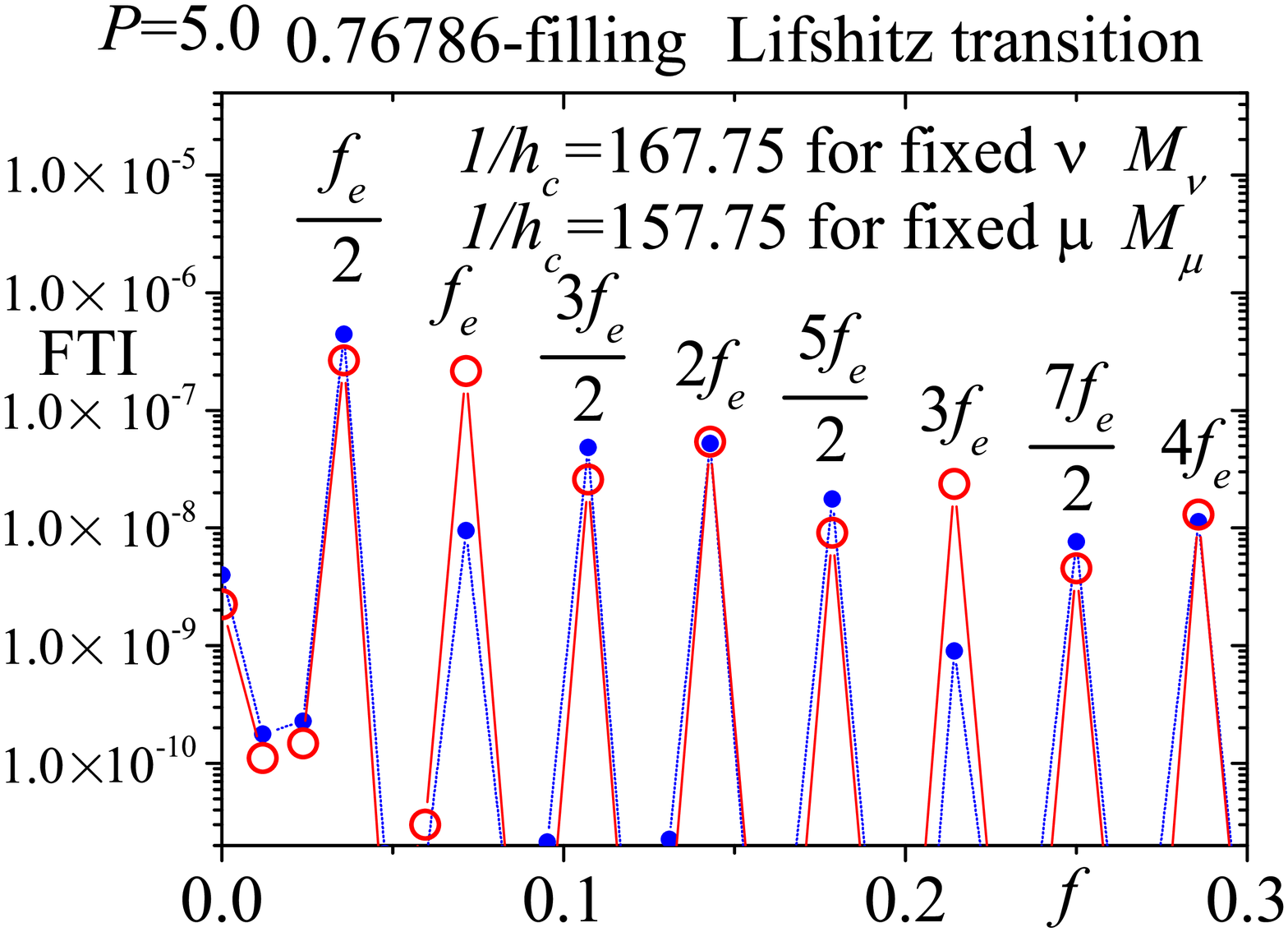}\vspace{-0.0cm}
\begin{flushleft} \hspace{0.5cm}(c) \end{flushleft}\vspace{-0.4cm}
\includegraphics[width=0.46\textwidth]{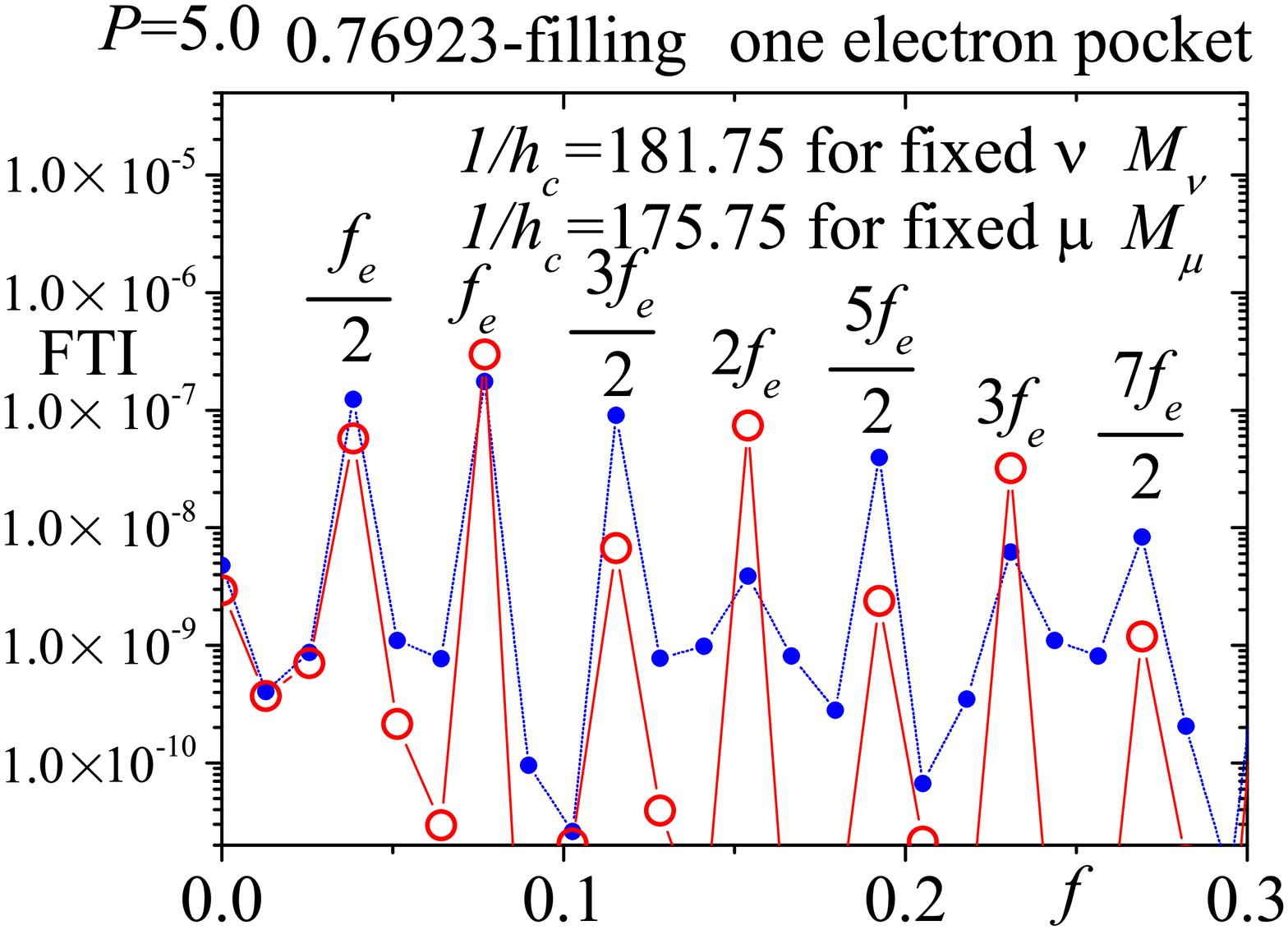}\vspace{-0.3cm}
\caption{
The FTIs of $M_{\nu}$ and $M_{\mu}$ at 0.76667-filling (a), at 0.76786-filling 
(b) and at 0.76923-filling (c) which are shown by red circles and blue dots, 
respectively. 
In (a), the range of the Fourier transform is 
$2L=3\times(2/f)=90$ and $1/h_c=179.75$ in $M_{\nu}$ and 
$2L=3\times(2/f)=90$ and $1/h_c=166.25$ in $M_{\mu}$. 
In (b), $2L=3\times(2/f)=84$ and 
$1/h_c=167.75$ in $M_{\nu}$ and $2L=3\times(2/f)=84$ and $1/h_c=157.75$ in $M_{\mu}$. In (c), 
$2L=3\times(2/f)=78$ and $1/h_c=181.75$ in $M_{\nu}$ and $2L=3\times(2/f)=78$ and $1/h_c=175.75$ in $M_{\mu}$.
}
\label{fig36_2}
\end{figure}

The magnetizations are not perfectly periodic near the Lifshitz transition, as shown in Figs. \ref{fig34} and \ref{fig35}. Nevertheless, we perform the Fourier transform in the finite range of $2L=3\times(2/f_e)$ at the center, $1/h_c$, and examine amplitudes of the Fourier components in $M_{\nu}$ and $M_{\mu}$ as a function of $1/h$. 
The Fourier transform of $M_{\nu}$ and $M_{\mu}$ are explained in Appendix \ref{AppendixD}. The Fourier transform intensities (FTIs) of $M_{\nu}$ and $M_{\mu}$ are shown in Figs.~\ref{fig36_2} (a), (b), and (c). 
There are large peaks at $f_e/2, f_e, 3f_e/2, 2f_e, \cdots$, where $f_e\simeq 0.0667$ at 0.76667-filling, $f_e\simeq 0.0714$ at 0.76786-filling, and $f_e\simeq 0.0769$ at 0.76923-filling, respectively.  
The frequency, $f_e$, is almost corresponding to the sum of the area of two small electron pockets at 0.76667-filling and 0.76786-filling ($A_e/A_{\rm BZ}\simeq 0.0668$ and $A_e/A_{\rm BZ}\simeq 0.0716$) and the area of one large electron pocket at 0.76923-filling ($A_e/A_{\rm BZ}\simeq 0.0768$). 
These frequencies are the same as those expected by the LK formula. 
However, the peaks at $f_e/2$ in $M_{\nu}$ and $M_{\mu}$ in Fig.~\ref{fig36_2} (c) are not expected by the LK formula, because there is only one electron pocket in Fig. \ref{fig8_0} (c). In the network model\cite{Pippard62,Falicov66,blount}, it is considered as an electron's effective closed orbital motion with the area, $A_e/2$, by the tunneling thorough the narrow neck in the presence of a magnetic field. Similarly, it is understood that the peaks at $3f_e/2$, $5f_e/2$, and $7f_e/2$ are due to the tunneling. 



From Figs. \ref{fig34} (b) and \ref{fig35} (b) (at the Lifshitz transition point), we can see that the magnetizations are almost periodic as a function of $1/h$. It can be understood from the above mentioned fact that the magnitudes of tunneling barrier on the Fermi surface is almost zero, as shown in Fig. \ref{fig8_0} (b).

\begin{figure}[bt]
\begin{flushleft} \hspace{0.5cm}(a) \end{flushleft}\vspace{-0.3cm}
\includegraphics[width=0.48\textwidth]{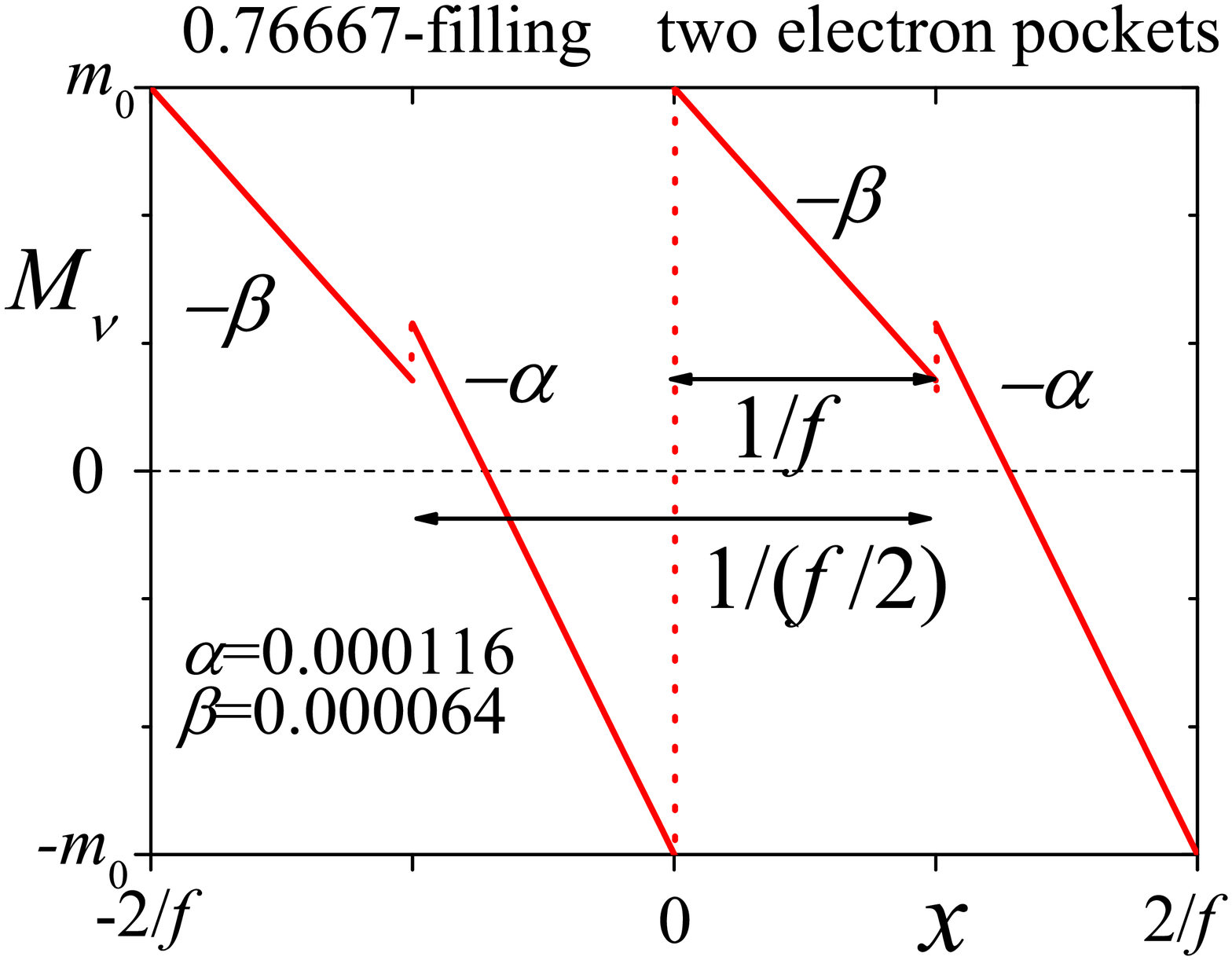}\vspace{-0.0cm}
\begin{flushleft} \hspace{0.5cm}(b) \end{flushleft}\vspace{-0.3cm}
\includegraphics[width=0.48\textwidth]{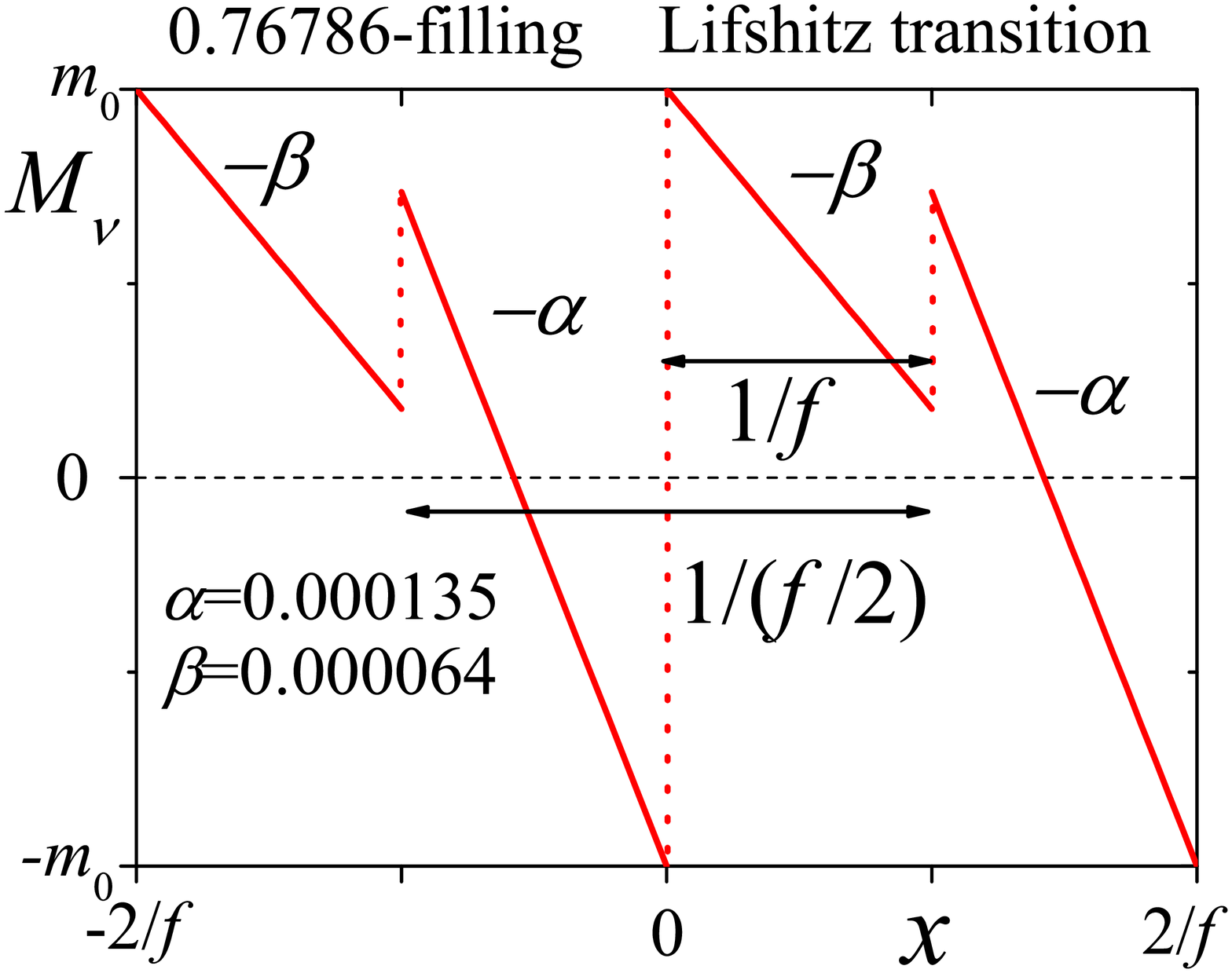}\vspace{-0.0cm}
\begin{flushleft} \hspace{0.5cm}(c) \end{flushleft}\vspace{-0.3cm}
\includegraphics[width=0.48\textwidth]{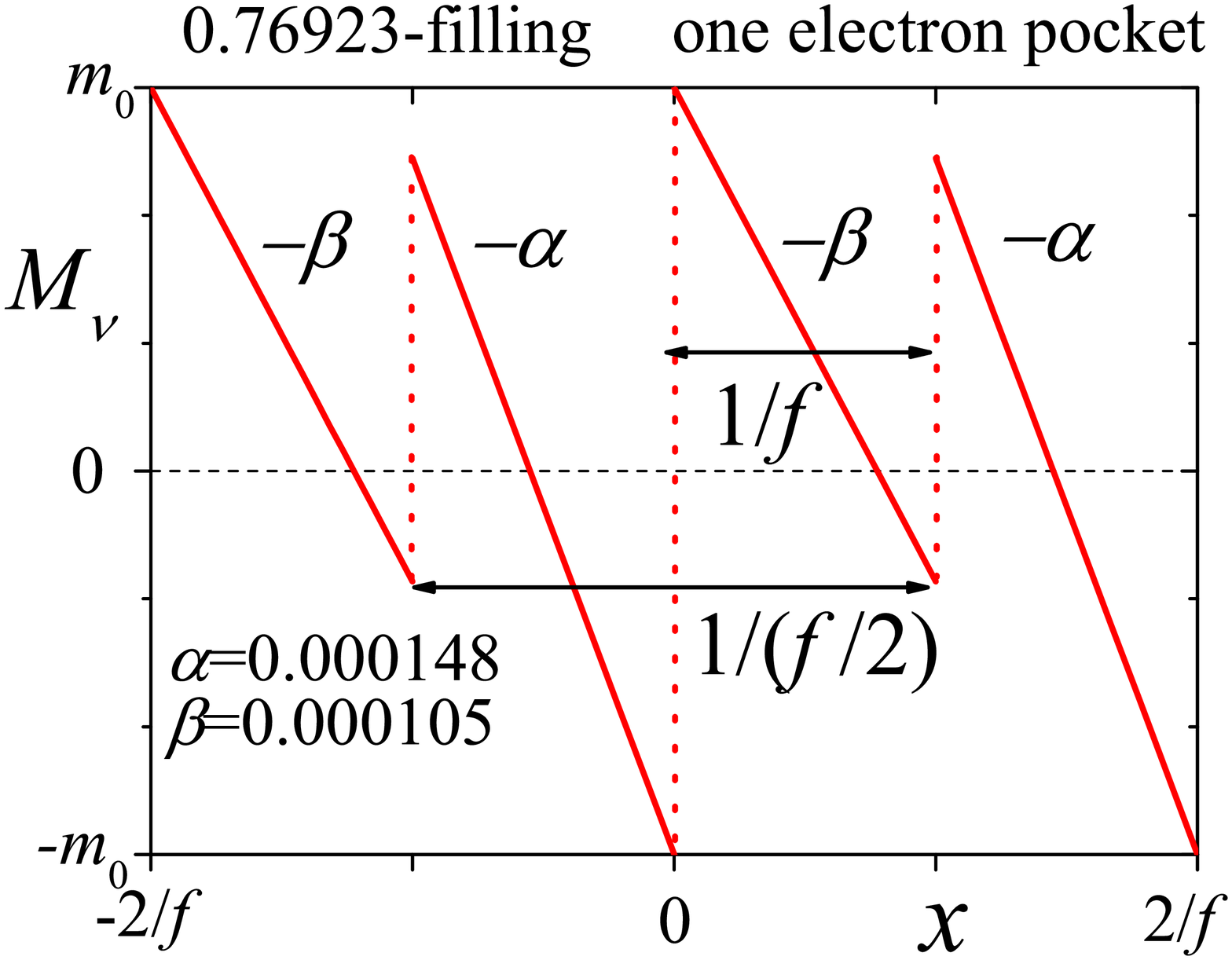}\vspace{-0.0cm}
\caption{
$M_{\nu}$ drawn by Eq. (\ref{f7}), where we use $\alpha=0.000116$ and $\beta=0.000064$ in (a), $\alpha=0.000135$ and $\beta=0.000064$ in (b), and $\alpha=0.000148$ and $\beta=0.000105$ in (c). We determine $\alpha$ and $\beta$ from Figs. \ref{fig34} (a), \ref{fig35} (a), and \ref{fig36} (a), respectively.
}
\label{fig_20}
\end{figure}

\begin{figure}[bt]
\begin{flushleft} \hspace{0.5cm}(a) \end{flushleft}\vspace{-0.2cm}
\includegraphics[width=0.48\textwidth]{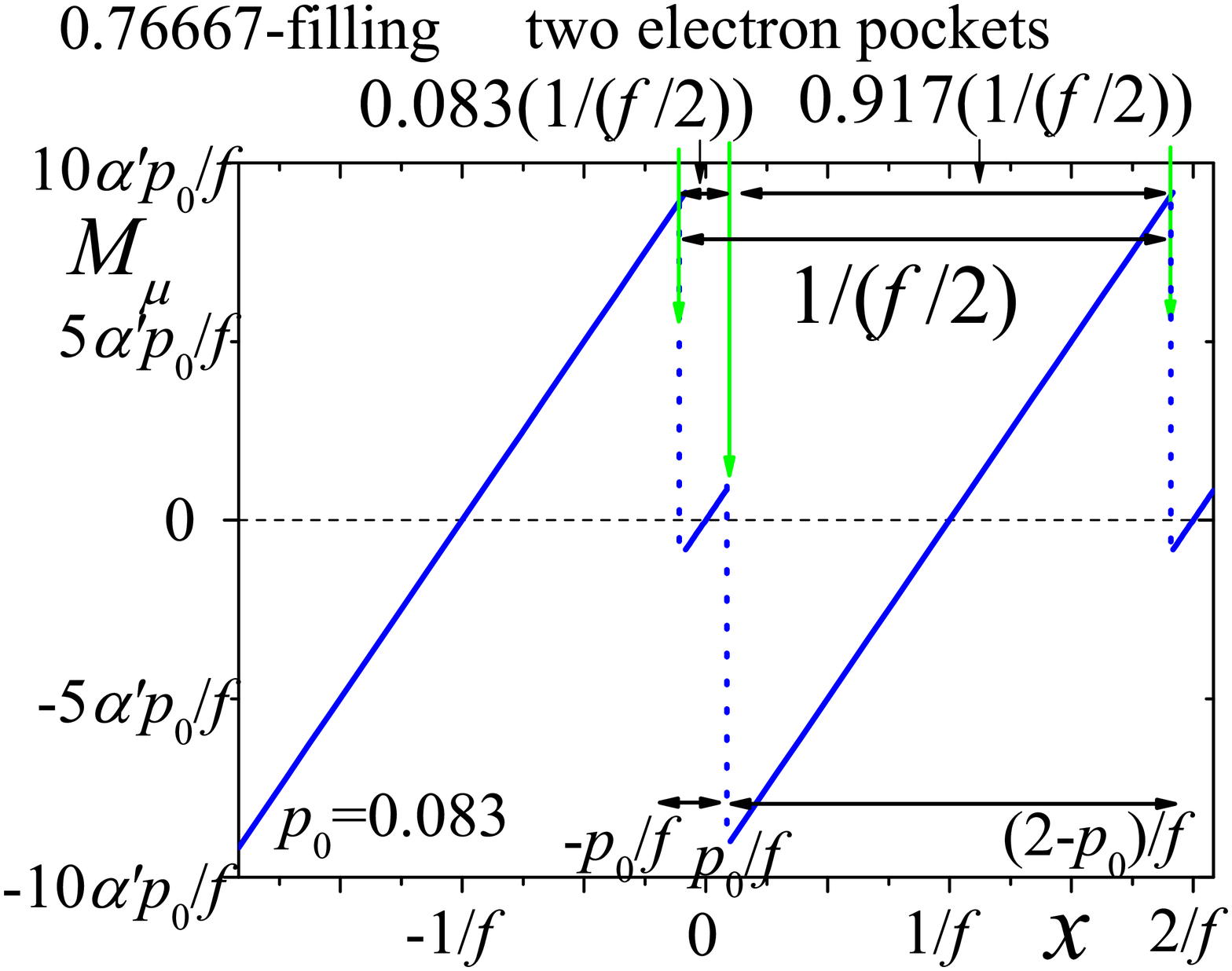}\vspace{-0.0cm}
\begin{flushleft} \hspace{0.5cm}(b) \end{flushleft}\vspace{-0.2cm}
\includegraphics[width=0.48\textwidth]{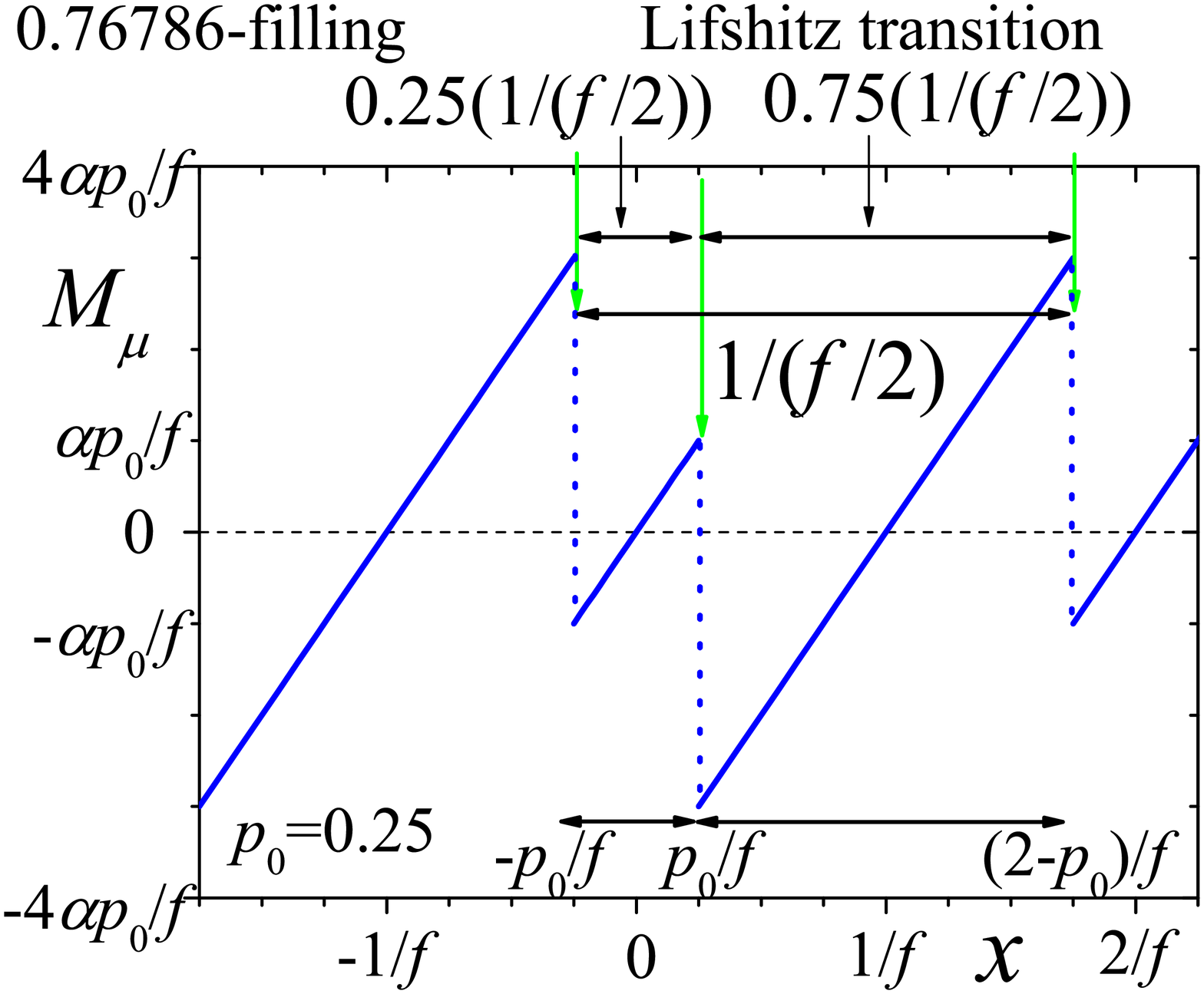}\vspace{-0.0cm}
\begin{flushleft} \hspace{0.5cm}(c) \end{flushleft}\vspace{-0.2cm}
\includegraphics[width=0.48\textwidth]{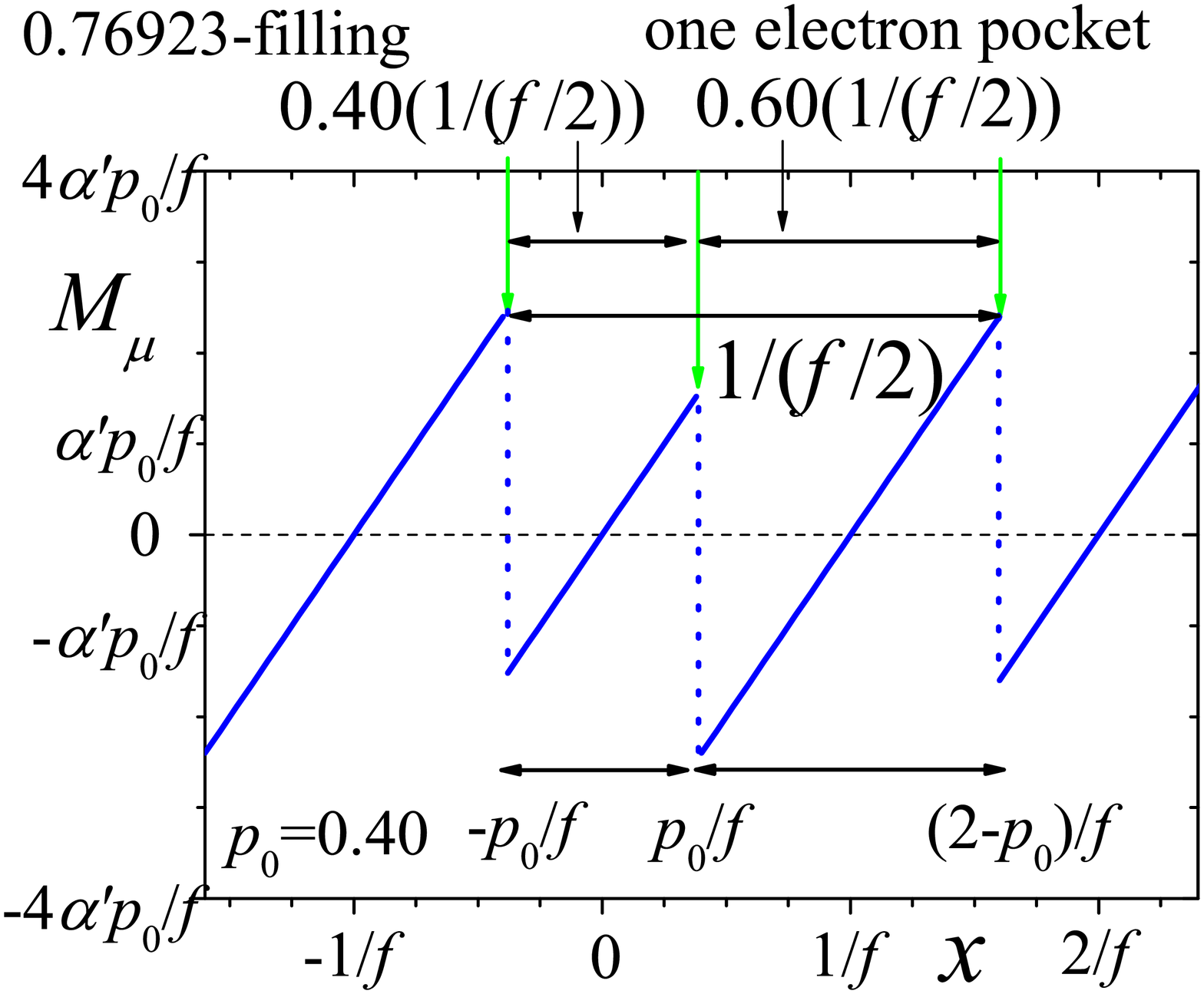}\vspace{-0.0cm}
\caption{
$M_{\mu}$ drawn by Eq. (\ref{f11_2}), where we use $p_0=0.1$ in (a), $p_0=0.25$ in (b) and $p_0=0.40$ in (c). We determine $p_0$ from Figs. \ref{fig34} (b), \ref{fig35} (b), and \ref{fig36} (b), respectively.}
\label{fig_19}
\end{figure}

To obtain the similar wave forms of $M_{\nu}$ and $M_{\mu}$ in Figs. \ref{fig34}, \ref{fig35}, and \ref{fig36}, 
we propose Eq. (\ref{Mod_saw}) with Eqs. (\ref{Mod_saw_a0}), (\ref{Mod_saw_al}) and (\ref{Mod_saw_bl}) for $M_{\nu}$ and Eq. (\ref{Mod_saw}) with Eq. (\ref{bl}) for $M_{\mu}$, respectively, which are explained in Appnedix \ref{AppendixE_0}. 
The wave forms by these equations are shown in Figs. \ref{fig_20} and \ref{fig_19}, which express the wave forms of Figs. \ref{fig34} (a), \ref{fig35} (a), and \ref{fig36} (a) and Figs. \ref{fig34} (b), \ref{fig35} (b), and \ref{fig36} (b), respectively. 

\begin{figure}[bt]
\begin{flushleft} \hspace{0.5cm}(a)
 \end{flushleft}\vspace{-0.3cm}
\includegraphics[width=0.5\textwidth]{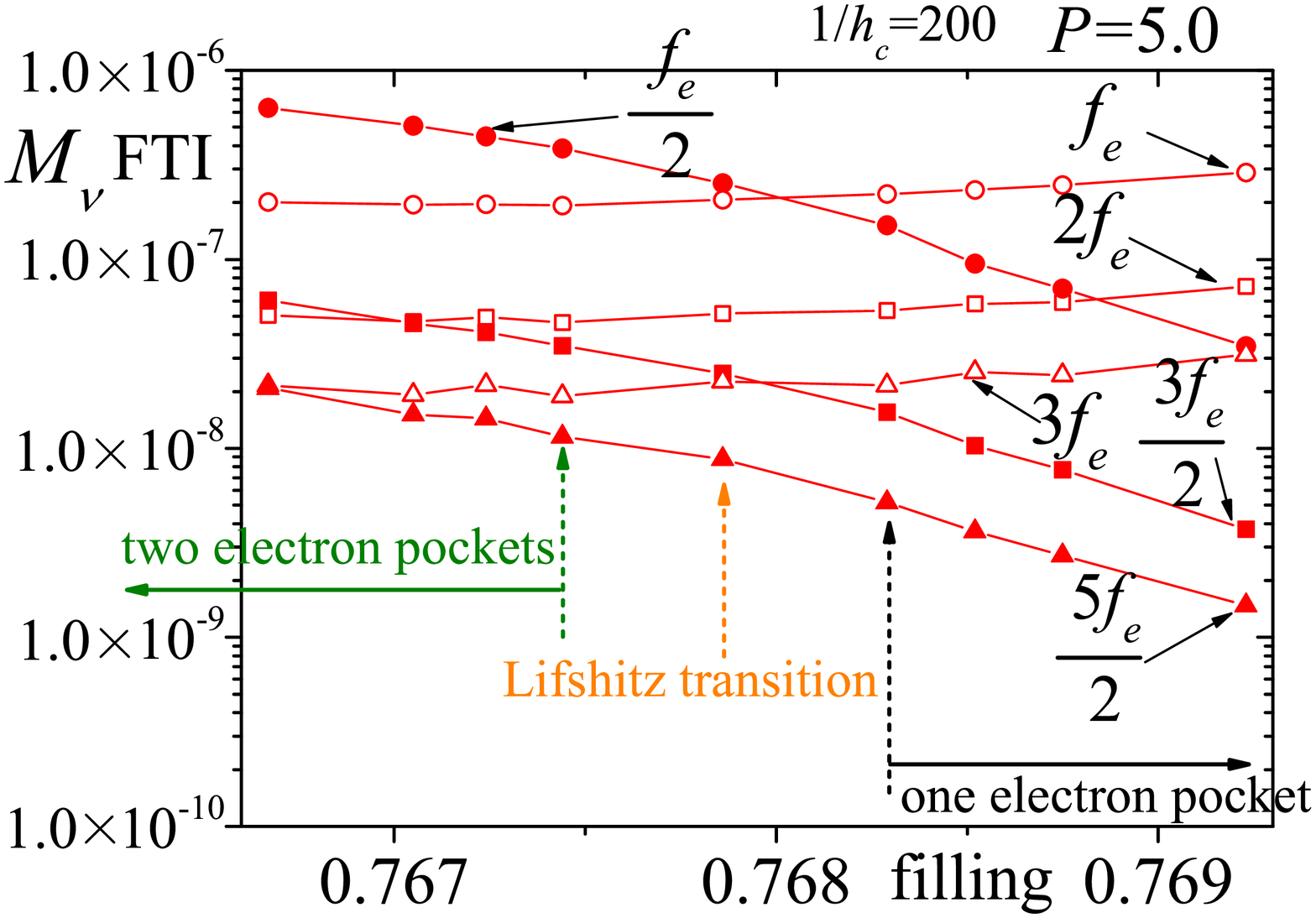}\vspace{-0.1cm}
\begin{flushleft} \hspace{0.5cm}(b) \end{flushleft}\vspace{-0.3cm}
\includegraphics[width=0.5\textwidth]{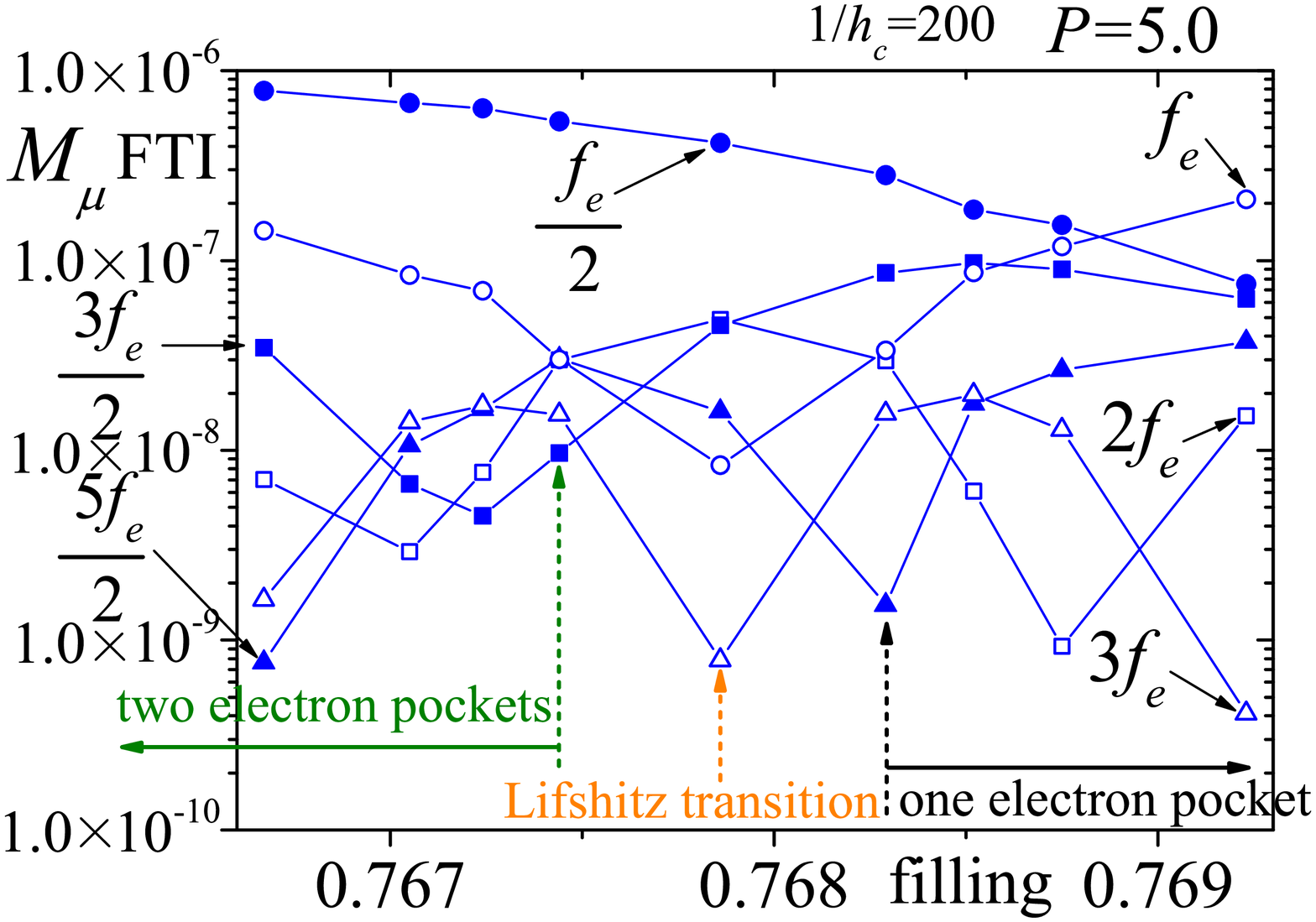}\vspace{0.1cm}
\caption{
The filling-dependences of the FTIs at $P=5.0$, where the range of the Fourier transform is $2L=4\times(2/f_e)=120, 118, 116, 114, 112, 110, 108, 106, 104$, and $1/h_c=200$ 
in 0.76667 0.76705, 0.76724, 0.76744, 0.76786, 0.76829, 0.76852, and 0.76923-fillings.}
\label{fig42}
\end{figure}

Next, we consider the filling-dependences of the FTIs, where the finite range 
of the Fourier transform is $2L=4\times(2/f_e)$. The filling-dependences of the FTIs at $f_e/2, f_e, 3f_e/2, 2f_e, 5f_e/2$, and 3$f_e$ in $M_{\nu}$ and $M_{\mu}$ are shown in Figs. \ref{fig42} (a) and (b), respectively. 
The filling-dependences of the FTIs in $M_{\nu}$ are quite different from those in $M_{\mu}$. Note that at the Lifshitz transition (0.76786-filling) 
the 3/2 and 5/2 times frequencies in $M_{\nu}$ are not enhanced in this system, although in the compensated metal these frequencies are enhanced\cite{KH2019}.



\section{Conclusions}

We have calculated the dHvA oscillations in the two-dimensional system with Dirac cones at $T=0$ numerically. By increasing the electron filling, $\nu$, the Lifshitz transition occurs from two electron pockets to one electron pocket. Since we ignore the effect of the spin, the dHvA oscillations in this study are caused by the Landau levels due to an electron's orbital motion. 
When two electron pockets with the same area exist at a far distance or there is only one electron pocket with a not narrow neck, the wave forms of the dHvA oscillations in $M_{\nu}$ (under the condition of the fixed electron filling) and $M_{\mu}$ (under the condition of the fixed chemical potential) are almost simple saw-tooth and these are inverted each other. These properties have been known well.

Near the Lifshitz transition of Fig. \ref{fig8_0}, however, we 
find very interesting features in the wave forms of the dHvA oscillations in $M_{\nu}$ and $M_{\mu}$. The additional center jump in the fundamental period of $2/f_e$ exists in $M_{\nu}$ and its jump always locates at the center in the fundamental period even when the magnetic field and the filling are changed. 
The center jump is larger as the electron filling increases. On the other hand, in $M_{\mu}$ a jump is separated into a pair of jumps and its separation position varies continuously as the electron filling is changed. These are caused by the lifting of doubly degenerated Landau levels which comes from the field-induced quantum tunneling. These phenomena in the dHvA oscillations have never been known. 
We propose the model which shows the similar dHvA oscillations near the Lifshitz transition [Eq. (\ref{Mod_saw}) with Eqs. (\ref{Mod_saw_a0}), (\ref{Mod_saw_al}) and (\ref{Mod_saw_bl}) for $M_{\nu}$ and Eq. (\ref{Mod_saw}) with Eq. (\ref{bl}) for $M_{\mu}$], respectively.

At the Lifshitz transition point, we also find some features. 
The 3/2 times and 5/2 times frequencies in $M_{\nu}$ are not enhanced, although the enhancements are seen in the compensated metal\cite{KH2019}. 
The dHvA oscillations in $M_{\nu}$ and $M_{\mu}$ are almost periodic as a function of the inverse of the magnetic field because of the smallness of tunneling barrier.

In this study, from the numerical approaches we make clear the influence which the lifting of the double degenerate Landau levels gives to the dHvA oscillations as a first step. In the next step, the Landau levels and the condition of the appearance of that lifting have to be obtained (for example, by the WKB approximation when the potential barrier is high, i.e., when the system is far from the Lifshitz transition), because these are useful to the study the dHvA oscillations near the Lifshitz transition in two-dimensional Dirac fermion systems (for example, we may predict the magnetic field strength when the additional center jump in $M_{\nu}$ begins to appear). However, it may not be easy to obtain these in the systems with the field-induced quantum tunneling, since the Landau levels in semi-Dirac\cite{Dietl2008} and in three-quarter Dirac\cite{HK2019}, where there is no the field-induced quantum tunneling, have been obtained from the careful studies. Therefore, these will be studied in future. 

If the spin of an electron is considered, the Landau levels are separated due to the Zeeman term. That separation becomes larger as the magnetic field increases. By the Zeeman term, it is expected that an additional center jump appears in $M_{\nu}$ and a jump is separated into a pair of jumps in $M_{\mu}$ in free electron model\cite{KHarxiv}. However, 
when the spacing of the Landau levels is much larger than that of the spin-splitting, the effect of spin can be ignored. In this paper, we perform the study in the simple spinless case as the first step. 
Since the effect of spin is interesting, the study including the spin-splitting in the lifting of the doubly degenerated Landau levels is needed in future.

Since the results in this paper are obtained in the ideal conditions ($T=0$, spinless, and no impurity), it is difficult to confirm experimentally our results in $\alpha$-(BEDT-TTF)$_2$I$_3$ under the doping and finite temperatures, where the saw-tooth wave form becomes broadening due to the thermal broadening\cite{shoenberg}, impurity effect\cite{shoenberg}, and so on. It has been known that the disorder can diminish the distinction between the wave form in the canonical and grand canonical ensembles\cite{Its2000}. 
Furthermore, the effect of the spin-splitting may appear. 
On the other hand, in the doping two-dimensional Dirac fermion systems with the small effective mass we expect that the continuous changes of the magnitude of the center jump in Fig. \ref{fig_20} and 
the spacing of the separated jump in Fig. \ref{fig_19} upon varying dopings may be observed qualitatively.

Although the calculations in the paper are performed in $\alpha$-(BEDT-TTF)$_2$I$_3$ under the doping, we expect that 
the obtained results will be widely observed in the system, for example, such as the doped graphene under the uniaxial 
strain\cite{Hasegawa2006,Rosenzweig}, black phosphorus\cite{Kim_2}, and 
twisted bilayer graphene\cite{lopes,aya}, where two Dirac points exist near each other in the momentum space. 
This is because the results provided by this study are attributed to only the lifting of doubly degenerated Landau levels and the similar lifting appears in that system (for example, it is clearly seen in energies under the magnetic field in the graphene with the anisotropic transfer integral\cite{HK2006}). 

\section*{Acknowledgement}
One of the authors (KK) thanks Naoya Tajima for useful discussions and informations of experiments.

\clearpage
\appendix

\section{Lifshitz and Kosevich formula}
\label{LKformula_dHvA}


Although 
the magnetizations should be calculated under the condition of the fixed electron number or the fixed electron filling, $\nu$, (canonical ensemble), Lifshitz and Kosevich\cite{shoenberg,LK} have calculated it under the condition of the fixed chemical potential, $\mu$,  (grand canonical ensemble). This is because the calculation in the grand canonical ensemble is justified if $\mu$ depends on the magnetic field less.

They have derived the LK formula\cite{shoenberg,LK} for the free electron model by using the semiclassical quantization rule\cite{Onsager}. Recently, it has been also shown that the LK formula can be used for the Dirac fermions\cite{Igor2004PRL,Igor2011,Sharapov}. The LK formula at $T=0$ for the two-dimensional multi closed Fermi surface with the area $(A_i)$ is given by
\begin{eqnarray}
M^{\rm LK}&=&M_0\sum_{i}
\widetilde{A}_i\sum_{l=1}^{\infty}\frac{1}{l}\sin\left[2\pi l\left(\frac{F_i}{H}-\gamma_i\right)
\right],\label{LK_0} \\
M_0&=&-\frac{e}{2\pi^2 c\hbar},\\
\widetilde{A}_i&=&\frac{|A_i|}{\frac{\partial A_i(\varepsilon^0)}{\partial \varepsilon^0}\big|_{\varepsilon^0=\mu}}, \label{Ai}
\end{eqnarray} 
where $i$ is the index for the closed orbit and the frequency ($F_i$) is given by 
\begin{equation}
F_i=\frac{c \hbar |A_i|}{2 \pi e}, 
\end{equation}
where 
$A_i>0$ and $A_i<0$ are for the electron pocket 
and for the hole pocket, respectively. In this paper, we consider only the case where one electron pocket or two electron pockets exist. 
When we use $h$ instead of $H$ in Eq. (\ref{LK_0}), we get 
\begin{equation}
\frac{F_i}{H}=\frac{f_i^{}}{h}, \label{C5}
\end{equation}
where 
\begin{equation}
f_i=\frac{|A_i|}{A_{\rm BZ}}. \label{C6}
\end{equation}
In Eq. (\ref{LK_0}), $\gamma_i$ is the phase of the oscillation, which comes from the phase of the Landau levels. 
When we consider the case of one electron pocket with the area of $A_e$ and the phase $\gamma_e$, Eq. (\ref{LK_0}) becomes 
\begin{eqnarray}
M^{\rm LK}&=&M_0
\widetilde{A}_e\sum_{l=1}^{\infty}\frac{1}{l}\sin\left[2\pi l\left(\frac{f_e}{h}-\gamma_e \right)
\right].\label{LK_02}
\end{eqnarray}

\begin{figure}[bt]
\begin{flushleft} \hspace{0.5cm}(a)
 \end{flushleft}\vspace{-0.3cm}
\includegraphics[width=0.29\textwidth]{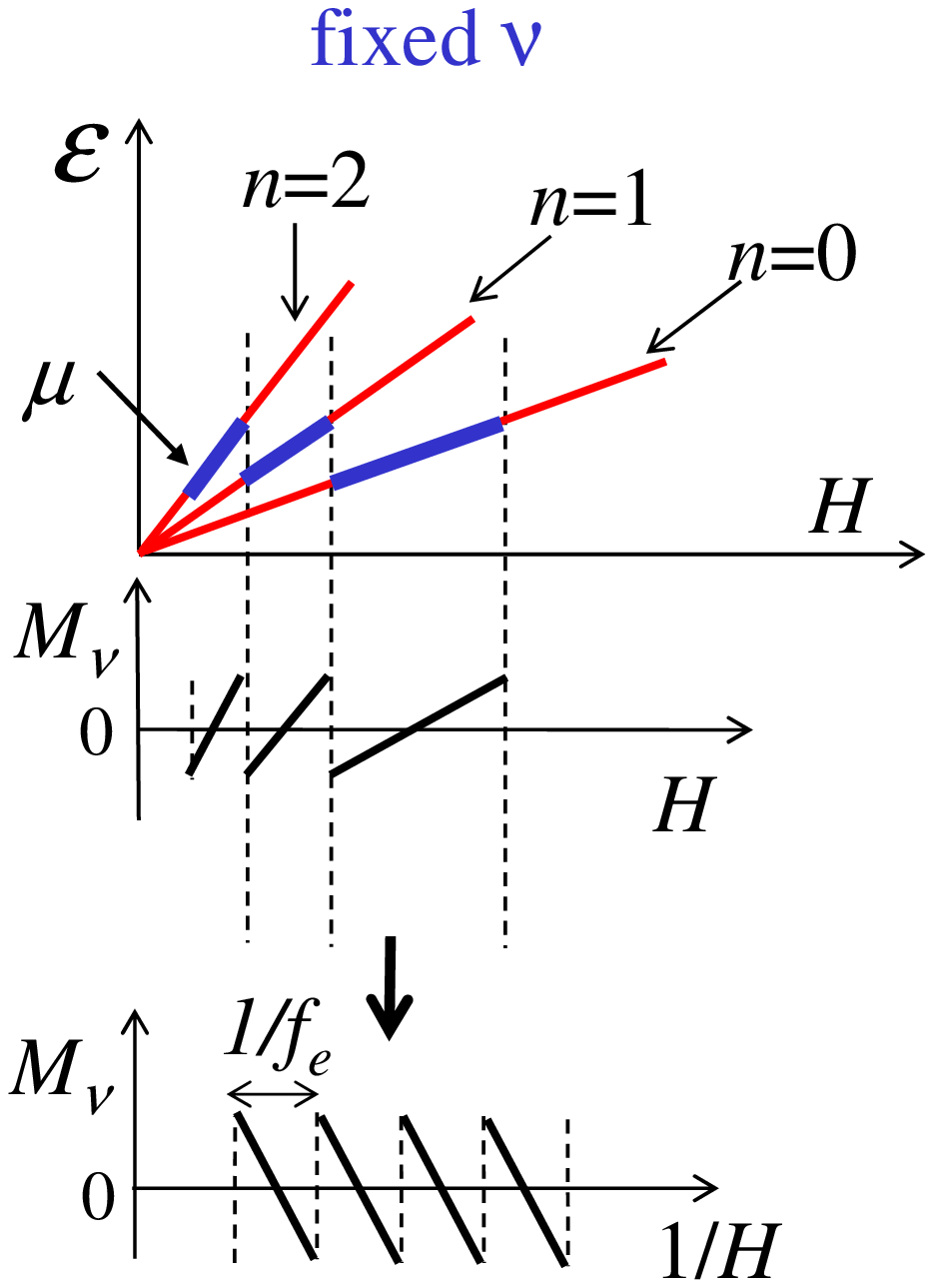}\vspace{-0.1cm}
\begin{flushleft} \hspace{0.5cm}(b) \end{flushleft}\vspace{-0.3cm}
\includegraphics[width=0.29\textwidth]{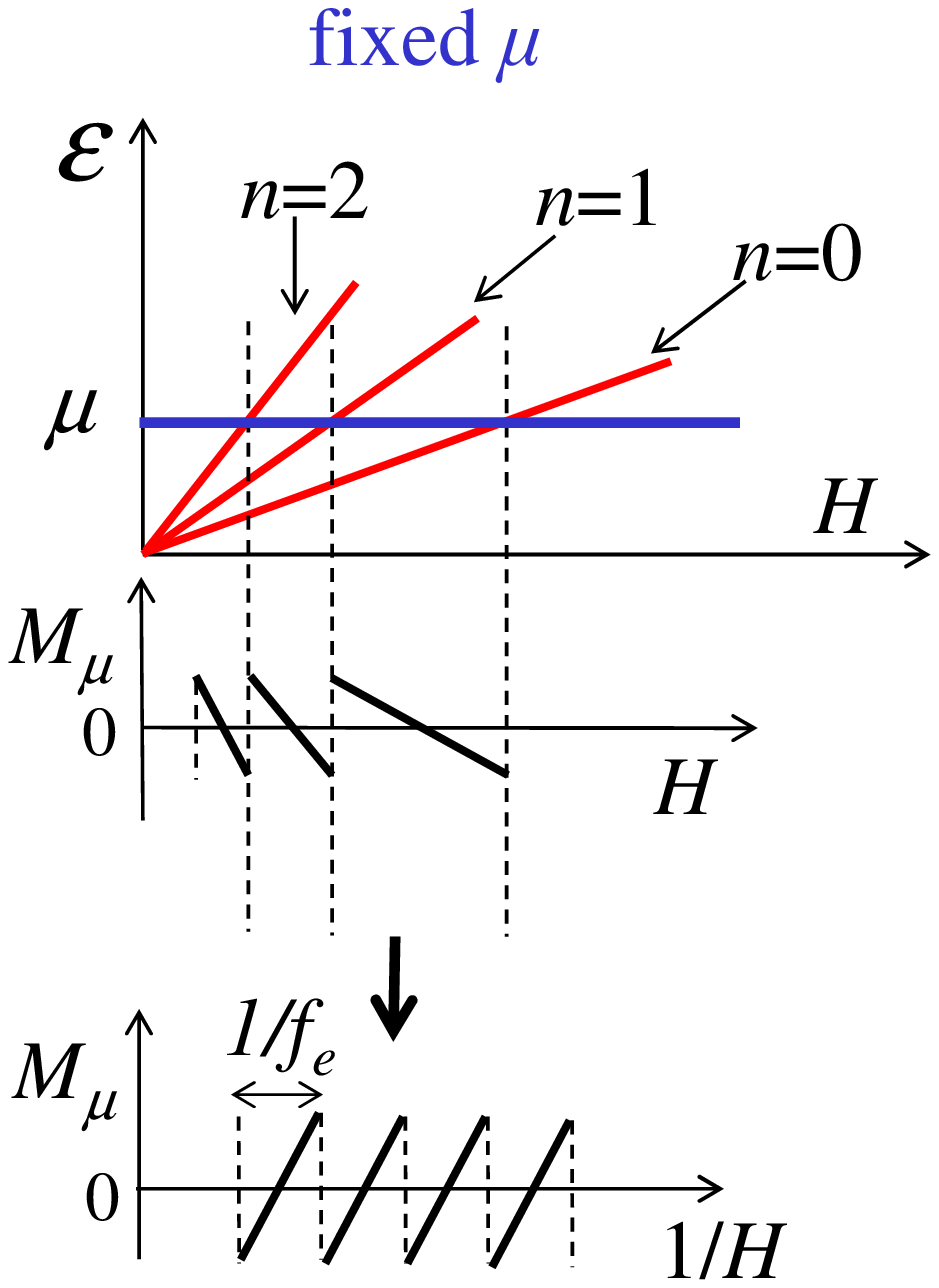}\vspace{0.1cm}
\caption{
In the two-dimensional system with a free electron pocket, schematic figures of the Landau levels (red lines), the chemical potential (blue lines) and the dHvA oscillations (black lines). (a) and (b) are for the conditions of 
the fixed electron number and of the fixed chemical potential, respectively. 
}
\label{fig37}
\end{figure}

Under the condition of the fixed $\nu$, the highest Landau level is partially filled, i.e., $\mu$ is pinned at the Landau level at $T=0$, as shown in Fig. \ref{fig37}(a), where the Landau levels for a free electron pocket are used. As the magnetic field is increased, the degeneracy of each Landau level increases, and $\mu$ jumps periodically as a function of $1/H$. These jumps are the origins of the dHvA oscillations, as shown in Fig. \ref{fig37}(a). Under the condition of the fixed $\mu$, the Landau levels and $\mu$ crosses periodically. Then, the dHvA oscillations appear, as shown in Fig. \ref{fig37}(b). The wave forms of the dHvA oscillations in $M_{\nu}$ and $M_\mu$ are the simple saw-tooth pattern and these are inverted each other. 

\section{semiclassical Landau quantization of energy}
\label{Appendix0}

When the uniform magnetic field is applied to perpendicular to the two-dimensional plane, the energies of the closed orbit in the wave-number space are quantized as $\varepsilon_n$ with integer $n$. In the semi-classical quantization rule\cite{Onsager}, the quantized energies are given by 
\begin{equation}
A(\varepsilon_n) = (n+\gamma) \frac{2 \pi e H}{\hbar c},
 \label{eqquantization0}
\end{equation}
where $A(\varepsilon_n)$ is the area of the closed orbit at $H=0$, $e$ is the electron charge, $c$ is the speed of light, $\hbar$ is the Planck constant divided by $2 \pi$ 
and $\gamma$ is a phase factor, which is related to the Berry phase\cite{berry,Wright2013}. To obtain the phase factor, the quantum mechanical calculations are needed.  For the free electron model and the Dirac fermions, $\gamma=1/2$ and $\gamma=0$ have been obtained, respectively. 

\section{Analytically derived Landau levels}
\label{Appendix01}

The Landau levels are analytically obtained as 
\begin{equation}
  \varepsilon_{n}^{\textrm{(free)}} \propto \left( n+\frac{1}{2} \right)  H, \ \ \ n=0, 1, 2
\label{eq2Dfree}
\end{equation}
for two-dimensional free electrons\cite{shoenberg,Landau} and 
\begin{equation}
\varepsilon_{n} ^{\textrm{(Dirac)}}\propto 
\pm\sqrt{|n| H},  \ \ \ n=0, 1, 2, \cdots \label{eq_1x}
\end{equation} 
for massless Dirac fermions (graphene\cite{Novo2005,McC1956} and 
$\alpha$-(BEDT-TTF)$_2$I$_3$\cite{Georbig2008,Morinari2009}, where the linearization of the energy dispersion has been done). 
The dispersion is quadratic along one axis (two directions, plus and minus directions) and
linear along one axis (two directions), when two Dirac points merge at a time-reversal invariant point\cite{Hasegawa2006}. That system is called the semi-Dirac system and the Landau levels are given by\cite{Dietl2008},
\begin{equation}
 \varepsilon_n^{\textrm{(semi-Dirac)}}\propto
\pm g(n)
 \left[\left(n+\frac{1}{2}\right)H\right]^{\frac{2}{3}}, \ n=0, 1, 2, \cdots, 
\label{eqmergedDirac}
\end{equation}
where $g(0) \simeq 0.808$, $g(\pm 1) \simeq 0.994$ and $g(n) \simeq 1$ for $|n| \geq 2$. 

Very recently, we have found that the Dirac cone is tilted critically at the critical pressure (2.3 kbar) in $\alpha$-(BEDT-TTF)$_2$I$_3$, where 
the linear term disappears and the quadratic term becomes dominant in one direction, while the linear term is finite in other three directions. This system is a three-quarter Dirac system\cite{KH2017}. We have also obtained that 
the Landau levels are given by\cite{KH2017,HK2019} 
\begin{equation}
 \varepsilon_n^{\textrm{(tq-Dirac)}}\propto
(nH)^{\frac{4}{5}}, \ n=0, \pm1, \pm2, \cdots. 
\label{tqDirac}
\end{equation}
Furthermore, when the Dirac cone is tilted horizontally such as the type III Weyl semimetal\cite{Hang2018}, we have shown that 
the Landau levels are given by\cite{KH2017} 
\begin{equation}
 \varepsilon_n^{\textrm{(type III)}}\propto
[(n+1)H]^2, \ n=0, 1, 2, \cdots. 
\label{type III}
\end{equation}

\section{total energies and magnetizations}
\label{appendix_m}

At $T=0$, the total energy ($E_{\nu}$) under the condition of the fixed electron number [i.e., the fixed electron filling ($\nu$)] is calculated by 
\begin{equation}
E_{\nu}=\frac{1}{4qN_k}\sum_{i=1}^{4qN_k\nu}
\varepsilon(i,{\bf k}), \label{E_nu}
\end{equation}
where $N_k$ is the number of $\mathbf{k}$ points taken in the magnetic Brillouin zone. 
In this system, $\nu=3/4$. 

The total energy ($E_{\mu}$) under the condition of fixed $\mu$ is calculated by 
\begin{equation}
E_{\mu}=\frac{1}{4qN_k}\sum_{\varepsilon(i,{\bf k})\leq\mu}  (\varepsilon(i,{\bf k})-\mu), \label{E_mu}
\end{equation} 
where $4q$ is the number of bands in the presence of magnetic field, and $\varepsilon(i,{\bf k})$ 
is the eigenvalues of $4q \times 4q$ matrix. 
The fixed chemical potential in Eq. (\ref{E_mu}) is given by 
\begin{equation}
\mu=\varepsilon_{\rm F}^0, 
\end{equation}
where $\varepsilon_{\rm F}^0$ is the Fermi energy at $h=0$.

The magnetizations for fixed $\nu$ and fixed $\mu$ are numerically calculated by 
\begin{eqnarray}
M_{\nu}&=& -\frac{\partial E_{\nu}}{\partial h}, \\
M_{\mu}&=& -\frac{\partial E_{\mu}}{\partial h},
\end{eqnarray}
respectively. 
If the $h$-dependence of $\mu$ is negligibly small, 
we obtain 
\begin{eqnarray}
M_{\nu}=M_{\mu}.
\end{eqnarray}



\section{Fourier transform intensities}
\label{AppendixD}
In order to analyze the oscillations in the magnetizations, 
we calculate the Fourier transform intensities numerically  as follows.
By choosing the center ($h_c$) and the finite range ($2L$), we calculate 
\begin{equation}
  \mathrm{FTI}(f, \frac{1}{h_c},L) =\left|\frac{1}{2L} 
  \int_{\frac{1}{h_c}-L}^{\frac{1}{h_c}+L} M (h) e^{2\pi i \frac{f}{h}} d \left(\frac{1}{h}\right)\right|^2, \label{FTI}
\end{equation}
where we take $f=j/(2L)$ with integer $j$ ($0\leq j\leq 256$ is used in this study). We take the finite range as 
$2L=3\times (2/f_e)$ in Figs. ~\ref{fig36_2} 
and $2L=4\times (2/f_e)$ in Fig. ~\ref{fig42}, respectively. 


\section{Amplitudes of Fourier coefficients of the modified saw-tooth model}
\label{AppendixE_0}

The periodical function, $M(x)$, with frequency, $f/2$, is given by the Fourier series
as
\begin{equation}
 M(x) = \frac{a_0}{2}+\sum_{\ell=1}^{\infty} \bigg[ a_{\ell} 
\cos (2\pi \ell \frac{f}{2} x) + b_{\ell} 
\sin(2\pi \ell \frac{f}{2} x)\bigg], \label{Mod_saw}
\end{equation}
where $a_0$, $a_\ell$ and $b_\ell$ are 
the Fourier coefficients. The FTIs of $M(x)$ becomes
\begin{equation}
\left\{ 
 \begin{array}{ll}
\frac{a^2_0}{4}, & \mathrm{if } \ \ell=0 \\
a^2_\ell+ b^2_\ell, 
& \mathrm{if } \   \ell=1, 2, 3, \cdots.
\end{array}
 \right.
\end{equation}

The saw-tooth dependence with period $1/(f/2)$ of the magnetizations as a function of
$1/h =x$ is given by 
\begin{equation}
 M^0(x) = \alpha x  \hspace{1cm} \mathrm{if} -\frac{1}{f} < x < \frac{1}{f}
\end{equation}
and
\begin{equation}
 M^0(x) = M^0(x+ \frac{2m}{f}),  \hspace{1cm} m=\pm 1, \pm 2, \cdots. 
\end{equation}
where the coefficients, $a_{\ell}$ and $b_\ell$, are given by
\begin{align}
 a_0&=a_\ell=0, \\
 b_\ell&= -\frac{2\alpha}{\pi f\ell}\cos(\pi\ell).
\end{align}
The $h$-dependence of Eq. (\ref{Mod_saw}) corresponds to that of Eq. (\ref{LK_02}).

To obtain the similar wave form of $M_{\nu}$ in this paper [Figs. \ref{fig34} 
(a), \ref{fig35} (a), and \ref{fig36} (a)], 
we introduce the modified saw-tooth model as 
\begin{equation}
 M_{\nu} (x) = \left\{ 
 \begin{array}{ll}
-\alpha x-m_0, & \mathrm{if } \ -\frac{1}{f} < x< 0 \\
-\beta x+m_0, 
& \mathrm{if } \   0< x < \frac{1}{f} \label{f7}
\end{array}
 \right.
\end{equation}
where $\alpha>0$ and $\beta>0$. 
Then, we obtain the Fourier coefficients as 
\begin{equation}
a_0 = \frac{\alpha-\beta}{2f},\label{Mod_saw_a0}
\end{equation}
\begin{equation}
a_\ell = \frac{\alpha-\beta}{\pi^2f\ell^2}\bigg[\cos(\pi\ell)-1\bigg], \label{Mod_saw_al}
\end{equation}
and
\begin{equation}
 b_\ell = \frac{2fm_0+(\alpha+\beta-2fm_0)\cos(\pi\ell)}{\pi f\ell}.\label{Mod_saw_bl}
\end{equation}
Note that 
the equation for $M_{\nu}$ has $a_0$ and $a_{\ell}$ of the cosine coefficients which are not included in the LK formula. 

For $M_{\mu}$, we propose as the following modified saw-tooth model \begin{eqnarray}
 M_{\mu}(x) = \left\{ 
 \begin{array}{ll}
 \alpha^{\prime} x, & \mathrm{if } \ -\frac{p_0}{f} < x< \frac{p_0}{f} \label{f11_2}\\
 \alpha^{\prime} \left(x-\frac{1}{f} \right), 
& \mathrm{if } \   \frac{p_0}{f} 
< x < \frac{2-p_0}{f}. 
\end{array}
 \right.\label{E11}
\end{eqnarray}
When $p_0=0.083$, the fundamental period [$1/(f/2)$] is divided into 
$0.083(1/(f/2))$ and $0.917(1/(f/2))$, as shown in 
Fig. \ref{fig_19} (a). Similarly, in the cases of $p_0=0.25$ and $p_0=0.40$, the fundamental periods are divided into 
$0.25(1/(f/2))$ and $0.75(1/(f/2))$ [see Fig. \ref{fig_19} (b)]  and $0.40 (1/(f/2))$ and $0.60(1/(f/2))$ [see Fig. \ref{fig_19} (c)], respectively. These describe the similar wave forms of $M_{\mu}$ in Figs. \ref{fig34} (b), \ref{fig35} (b), and \ref{fig36} (b), respectively. 

From the Fourier transform of Eq. (\ref{E11}), we 
obtain the Fourier coefficients as 
\begin{equation}
a_0 = a_\ell =0 \label{al}
\end{equation}
and
\begin{equation}
 b_\ell = - \frac{2\alpha^{\prime}}{\pi f \ell} 
\cos(\pi\ell p_0).\label{bl}
\end{equation}





\end{document}